\documentclass[amssymb,amsmath,prd,twocolumn,preprintnumbers,superscriptaddress,nofootinbib]{revtex4-1}

\usepackage{graphicx}
\usepackage{bm}
\usepackage{amssymb,amsmath}
\usepackage{mathrsfs}
\usepackage{latexsym}
\usepackage{color}
\usepackage[normalem]{ulem} 
\usepackage{dcolumn}
\usepackage[colorlinks=true,citecolor=blue,urlcolor=blue]{hyperref}
\usepackage[usenames,dvipsnames]{xcolor}
\usepackage{xspace}
\usepackage{graphicx}

\usepackage{multirow}
\usepackage{xfp} 
\usepackage{comment}
\usepackage{nameref}
\usepackage{lineno}

\newcommand{\new}[1]{{#1}}
\newcommand{\fsn}{{\sc failed-sn}}
\newcommand{\plp}{{\sc powerlaw + peak}}
\newcommand{\bpl}{{\sc broken powerlaw}}
\newcommand{\fsnplp}{{\sc failed-sn + powerlaw + peak}}
\newcommand{\fsnbpl}{{\sc failed-sn+broken powerlaw}}
\newcommand{\default}{{\sc default}}
\newcommand{\pairing}{{\sc pairing}}

\newcommand{\chieff}{\chi_{\rm eff}}
\newcommand{\rateunits}{$M_{\odot}^{-1}\mathrm{yr}^{-1} \mathrm{Gpc}^{-3}$}

\usepackage[english]{babel}
\usepackage{xspace}
\usepackage{color,soul}

\newcommand{\result}[1]{#1}
\newcommand{\massresult}[1]{$#1\, M_{\odot}$}

\newcommand{\bplfracsourcesinpeak}{\result{$0.75^{+0.08}_{-0.16}$ }}

\newcommand{\bpluppermassminstar}{\massresult{13.46_{-6.77}^{+5.16}}}
\newcommand{\bplfailedmassmaxstar}{\massresult{11.93_{-0.87}^{+1.35}}}
\newcommand{\bpldeltamaxminstar}{\massresult{1.43_{-6.79}^{+5.27}}}

\newcommand{\bplfsnbumpcenter}{\massresult{9.95_{-1.25}^{+0.53}}}
\newcommand{\bplfsnbumpwidth}{\massresult{0.68_{-0.35}^{+0.80}}}
\newcommand{\bplfsnbumpmaxrate}{\result{$8.41_{-4.22}^{+6.30}$}}
\newcommand{\bplmaxrateindip}{\result{$0.19_{-0.16}^{+0.20}$}}

\newcommand{\bplprobmturnlessmmax}{\result{$0.61$}}

\newcommand{\plpfracsourcesinpeak}{\result{$0.76^{+0.07}_{-0.17}$ }}
\newcommand{\plpupperbeta}{\result{$1.00_{-2.04}^{+1.87}$}}
\newcommand{\plpfailedbeta}{\result{$-0.05_{-1.74}^{+1.81}$}}
\newcommand{\plpfailedqmin}{\result{$0.28_{-0.19}^{+0.22}$}}
\newcommand{\plpupperqmin}{\result{$0.28_{-0.20}^{+0.23}$}}

\newcommand{\plpuppermassminstar}{\massresult{16.21_{-6.91}^{+3.81}}}
\newcommand{\plpfailedmassmaxstar}{\massresult{12.06_{-0.93}^{+1.37}}}

\newcommand{\plpdeltamaxminstar}{\massresult{4.07_{-6.69}^{+4.00}}}

\newcommand{\plpfsnbumpcenter}{\massresult{9.95_{-1.25}^{+0.53}}}
\newcommand{\plpfsnbumpwidth}{\massresult{0.71_{-0.35}^{+0.75}}}
\newcommand{\plpfsnbumpmaxrate}{\result{$7.36_{-3.11}^{+6.35}$}}
\newcommand{\plpmaxrateindip}{\result{$0.24_{-0.24}^{+0.37}$}}

\newcommand{\plplowraterange}{\massresult{(12.0,16.1)}}
\newcommand{\bpllowraterange}{\massresult{(12.3,14.1)}}

\newcommand{\plpfracturnonabovemaxfailed}{\result{0.88}}

\newcommand{\plpoverbplbayesfactor}{\result{$\sim 2.6$}}
\newcommand{\plpoverbpllikelihoodratio}{\result{$\sim 20$}}
\newcommand{\plpoverdefaultbayesfactor}{\result{$\sim 0.21$}}

\newcommand{\plpoverdefaultlikelihoodratio}{\result{$\sim 547$}}
\newcommand{\plpmaxlikelihoodoverseparatepops}{\result{$ \sim 8.9$}}

\newcommand{\plpnolowmassfsnwidth}{\massresult{0.56_{-0.25}^{+0.65}}}

\newcommand{\plpnolowmassqmin}{\result{$0.46_{-0.25}^{+0.19}$}}

\newcommand{\plpforcednogapoverdefaultbayesfactor}{\result{$0.032$}}
\newcommand{\plpforcednogapoverdefaultmaxlratio}{\result{$3.6$}}

\newcommand{\fsnplpbumpchieffmean}{\result{$0.05_{-0.03}^{+0.03}$}}
\newcommand{\fsnplpbumpchieffsigma}{\result{$0.07_{-0.02}^{+0.03}$}}
\newcommand{\fsnplpupperchieffmean}{\result{$0.04_{-0.03}^{+0.03}$}}
\newcommand{\fsnplpupperchieffsigma}{\result{$0.13_{-0.03}^{+0.03}$}}

\newcommand{\plpprobmturnlessmmax}{\result{$0.88$}}

\newcommand{\plppowerlawindex}{\result{$3.57_{-0.73}^{+0.60}$}}

\newcommand{\defaultalphaone}{\result{$0.58_{-1.07}^{+0.73}$}}
\newcommand{\defaultalphatwo}{$5.82_{-0.82}^{+1.08}$}

\newcommand{\montecarlolimitlowraterange}{\result{(13.0, 15.2)}}

\newcommand{\CIT}{\affiliation{TAPIR, California Institute of Technology, Pasadena, CA 91125, USA}}
\newcommand{\CITLab}{\affiliation{LIGO Laboratory, California Institute of Technology, Pasadena, California 91125, USA}}

\newcommand{\UIUC}{\affiliation{Illinois Center for Advanced Studies of the Universe, Department of Physics, University of Illinois at Urbana-Champaign, \\Urbana, Illinois, 61801, USA}}

\begin{document}
\title{
Low-mass failed supernovae and the $10\,M_{\odot}$ peak in the merging black hole mass distribution
}

\author{Isaac Legred}\email{ilegred2@illinois.edu}\CIT \CITLab \UIUC
\author{Jacob Golomb}\email{jgolomb@caltech.edu}\CIT \CITLab
\author{Katerina Chatziioannou}\email{kchatziioannou@caltech.edu}\CIT \CITLab

\begin{abstract}
    Gravitational-wave observations reveal that the rate of merging black holes drops by $\sim2$ orders of magnitude from component masses $\sim 10\,M_{\odot}$ to $\sim 15\,M_{\odot}$.  
    The increased compactness of the black hole progenitor cores may contribute to the $\sim 10\,M_{\odot}$ overdensity, but cannot fully explain the rate difference.    
    In this paper, we consider the possibility that the overdensity is reinforced by supernova processes that result in efficient black hole formation from direct collapse in a narrow range around $10\, M_{\odot}$. 
    We extend previous studies by considering a distinct subpopulation of failed-supernovae black holes, possibly separated by a gap in the primary mass distribution from the rest of the population.   
    Using 153 observations from the latest GWTC-4.0 catalog, we confirm a strong peak in the primary mass distribution at $10\,M_{\odot}$, with a peak rate density of \plpfsnbumpmaxrate{}\,\rateunits{}. 
    The rate drops sharply and becomes consistent with zero at the 90\% level for primary mass $m_1\in$ \plplowraterange{}, then rises again to confidently nonzero values above $\sim 16\,M_{\odot}$ before falling at higher masses.
    Our results reveal structure in the mass distribution in the $10-20\,M_{\odot}$ range, with rate changes of multiple orders of magnitude across a few solar masses, consistent with a distinct population of failed-supernova black holes.
    \new{These conclusions persist in the latest GWTC-5.0 catalog.}
\end{abstract}

\maketitle

\section{Introduction}
\label{sec:introduction}

The $\sim 200$ gravitational-wave signals from merging binary black holes (BBHs)~\cite{LIGOScientific:2025slb} identified so far by the LIGO~\cite{LIGOScientific:2014pky}, Virgo~\cite{VIRGO:2014yos}, and KAGRA~\cite{KAGRA:2020tym} detectors inform the distribution of BH masses and spins across cosmic time~\cite{LIGOScientific:2025pvj}. 
The BBH properties carry imprints of both the process through which progenitor stars died to form the observed BHs and the environment in which the binaries formed and evolved.  
In isolated formation, a BBH system forms from the collapse of two massive stars, with binary interactions shrinking the orbit such that the resulting binary can merge under gravitational radiation~\cite{Peters:1963ux, Postnov:2014tza}.  
In dynamical formation, interactions in dense environments cause existing binaries to tighten or binaries to form by capture~\cite{PortegiesZwart:2000dm,Rodriguez:2016avt}.  
Theoretical predictions~\cite{Mapelli:2021taw} and observations~\cite{Belczynski:2022wky, Zevin:2020gbd} suggest that at least some BBHs form via the isolated channel (though see Ref.~\cite{Ye:2025ano}).
BBH data can probe stellar evolution and massive star collapse under either formation mechanism, coupled to further binary dynamical or stellar evolution processes.

The BBH mass distribution has a global maximum near $10\,M_{\odot}$, after which the rate density drops by 1-2 orders of magnitude~\cite{GWTC2pops, GWTC3pops, Edelman:2022ydv, Farah:2023vsc, Callister:2023tgi, Toubiana:2023egi, Gennari:2025nho, Bertheas:2026odj}. 
This conclusion is supported both by low-dimensional ``parametric" models (powerlaws, Gaussian peaks, dips, etc.)~\cite{GWTC3pops, LIGOScientific:2025pvj} and more flexible ``nonparametric" models (Gaussian processes, splines, etc.)~\cite{Edelman:2021zkw, Callister:2023tgi, Farah:2023vsc}.
Compared to more massive binaries, BBHs whose primary component falls in this range are characterized by a broad mass-ratio distribution and low spins~\cite{Li:2022jge,Godfrey:2023oxb,Galaudage:2024meo,LIGOScientific:2025pvj,Banagiri:2025dmy,Tong:2025xir}.
Furthermore, observations show a ``gap" in the distribution of the better-constrained binary chirp mass~\cite{Tiwari:2020otp,Tiwari:2025lit,Tiwari:2025oah} near $10-12\,M_{\odot}$.
This does not automatically imply a gap in the primary mass, which will depend on the details of currently poorly-constrained binary mass ratio distribution~\cite{Tiwari:2025lit}.
The potential sharp peak and subsequent gap would imply a separate astrophysical population that does not bleed over into the $12-16\,M_{\odot}$ range and a high mass population that does not extend below $\sim 16\,M_{\odot}$.

While data decisively favor a peak at $10\,M_{\odot}$ distinct from the global power-law~\cite{LIGOScientific:2025pvj}, evidence for a gap in the primary mass distribution is elusive.
Since the feature under discussion involves a rate reduction by 1-2 orders of magnitude within a few solar masses, modeling choices are important.
Nonparametric models attempt to dispense with modeling assumptions in favor of flexibility.
Nonetheless, they still inevitably impose correlation lengths on the mass distribution, for example via smoothing priors~\cite{Heinzel:2024jlc}. 
With the caveat that nonparametric models may not identify features sharper than their (explicit or implicit) correlation scale, multiple studies have found no compelling evidence of a gap~\cite{Edelman:2021zkw, Callister:2023tgi, Farah:2023vsc,LIGOScientific:2025pvj} in the primary mass distribution.
Parametric models, on the other hand, use functional forms to model the population, which is justified by \emph{a priori} expectations about the overall population.  
In this context, explicit parametric gap-like features were confidently identified in the chirp mass~\cite{Tiwari:2025lit}, but not the primary mass distribution~\cite{Adamcewicz:2024jkr}. 
A more detailed discussion of these works and comparison with our results is given in Sec.~\ref{sec:conclusions}. 

A potential explanation for the peak at $10\,M_{\odot}$ is linked to the properties of the BH progenitors, and specifically, the increased compactness of progenitor stellar cores which is related to their explodability (and hence the difficulty of forming BHs).
Stars with carbon-oxygen (CO) core mass $M_{\rm CO}\sim 6.5\,M_{\odot}$ have anomalously large core compactness which in turn may lead to direct collapse to BHs~\cite{Schneider:2023mxe}, a fact also established in phenomenological 1D simulations~\cite{OConnor:2010moj} (see also Ref.~\cite{Just:2018djz} and references therein). 
Such a mechanism would naturally explain peaks at $\sim 9\, M_{\odot}$ and $16\, M_{\odot}$, and a gap from $9-13\, M_{\odot}$~\cite{Schneider:2023mxe}.  
Moreover, BHs formed from direct collapse of overcompact massive stars should have lower kicks~\cite{Burrows:2024wqv}, increasing the chances of binary survival. 
A model directly based on this scenario is explored in Ref.~\cite{Galaudage:2024meo}, which however did not allow for an arbitrarily low-rate gap in the mass distribution.
However, more detailed population synthesis estimates suggest a rate enhancement of only a factor of $\sim 3$ at $\sim 9-10\,M_{\odot}$~\cite{Schneider:2023mxe, Willcox:2025cus,Willcox:2025poh}, far below the observed drop.

Instead, it is possible that failure of stars to launch a supernova (SNe) and instead collapse directly to BHs depends also on other details of the progenitor stars~\cite{Disberg:2023gel}.
3D simulations of dying massive stars have achieved 
successful explosions across a range of progenitor masses~\cite{Mezzacappa:2007ad, Vartanyan:2018iah, Shankar:2025pwq, Murphy:2025vyv, Powell:2018isq, Janka:2016fox, Pan:2020idl, Bugli:2021stj, Burrows:2024pur},  even with no tunable macroscopic calibration parameters.
However, expectations that explodability is primarily driven by the stellar core compactness~\cite{OConnor:2010moj} (which in turn is determined primarily by the ZAMS mass) have not been confirmed, with factors such as the compositional and entropy profiles coming into play~\cite{Boccioli:2022ktx}. 
Among the solar-metallicity progenitors of Ref.~\cite{Burrows:2024wqv}, only the ones with ZAMS mass $11-14\,M_{\odot}$ formed BHs due to failed SNe.  This is characteristic of a trend seen in Refs.~\cite{Couch:2019mrd, Sykes:2024mel} in which progenitors in this mass range are prone to fail to launch SNe.  
Varying the progenitor ZAMS mass, \citet{Burrows:2024pur} found progenitors near $12\,M_{\odot}$ to be the only ones failing to explode (though this depends on metallicity~\cite{Heger:2002by,Schneider:2023mxe, Maltsev:2025bgs}).
Stars of slightly lower and higher ZAMS masses instead exploded and left behind NSs or much lower mass BHs, see \emph{e.g.} Fig.~2 of Ref.~\cite{Burrows:2024wqv}.   
It may be the case that at solar metallicity, $\sim12-14\,M_{\odot}$ progenitors are uniquely vulnerable to direct collapse to BHs.

The above results point to the possibility that a unique isolated channel for forming BBH progenitor systems, that is largely inoperative at slightly higher or lower masses, may explain the ${\cal{O}}(1-2)$ order of magnitude rate enhancement of $\sim 10\, M_{\odot}$ BHs.     
Of course, BHs are observed at many different masses globally following power-law distributions~\cite{GWTC2pops, GWTC3pops}.
This background power law is potentially produced by physically distinct progenitors, such as higher-mass stars with lower metallicity.
If its turn-on occurs at a mass greater than the peak at $\sim10\, M_{\odot}$, then we expect a highly suppressed merger rate, or even a ``gap," between the turn-off of the failed SNe channel and the turn-on of the power law. 
While the existence of a gap is not guaranteed (since it depends on the properties of the background component or it could be filled by hierarchical mergers~\cite{Tong:2025xir,Farah:2026jlc}), identifying it would be strong evidence that the low-mass population is astrophysically distinct from the high-mass one.

In this study, we are motivated by these theoretical arguments to propose a modest extension of existing models, geared toward identifying order-of-magnitude rate changes over a few solar masses. 
Our model considers the possibility that all BBH mergers with primary components near $10~M_{\odot}$ represent a \emph{separate population} from the overall mass distribution.   
Expanding upon Ref.~\cite{Galaudage:2024meo}, we permit there to be a gap between the high-mass systems ($m_1 \gtrsim 15\, M_{\odot}$) and the peak at $10\, M_{\odot}$.  
This allows us to both model the physics of the ``peak," and also to target the turn-on of the remainder of the population. 
The model, described in Sec.~\ref{sec:failed-sn-models}, contains two components.
The failed-SN peak is modeled as a truncated Gaussian distribution on the primary mass while high-mass mergers are modeled with either a power-law and a Gaussian peak or a broken power-law.
The peak and background populations can have distinct mass ratio and spin distributions, thus allowing for different pairing within each population but no pairing ``between" them.
We assume that the two populations share a redshift distribution~\cite{GWTC2pops, GWTC3pops, LIGOScientific:2025pvj}.

Our results are presented in Sec.~\ref{sec:failed-sn-results}.
The fraction of BBHs with a primary in the failed-SN peak is \plpfracsourcesinpeak (median and $90\%$ symmetric credible regions), indicating that the majority of  BBHs are consistent with this subpopulation. 
The distribution of merging BBHs is further consistent with a separation between the $\sim10 M_{\odot}$ peak and the higher-mass component of the population, with the data preferring but not requiring a ``gap."
The turnover mass at which the high-mass population begins is \plpuppermassminstar. 
This is likely, but not certainly, higher than the typical primary in the failed-SN  component at $10\, M_{\odot}$. 
Since the data are consistent with a gap, the merger rate becomes consistent with zero at 90\% for primary masses in  \plplowraterange.\footnote{In practice, the 90\% lower limit is below $\sim10^{-3}$\,\rateunits{} which we do not consider to be meaningfully different from zero, see Sec.~\ref{sec:failed-sn-results}.}

We discuss implications of these results and compare them to previous work in Sec.~\ref{sec:conclusions}.
It is plausible from current data that most stellar-mass BBHs form from a specific failed-SN channel and that the background power law component which tracks the stellar mass function at different metallicities only turns on near $16\,M_{\odot}$~\cite{Schneider:2023mxe}.

\section{Failed Supernova Population Model}
\label{sec:failed-sn-models}

In this section, we describe the population model motivated by the potential for a narrow-mass subpopulation of ``failed-SN."
We focus on the salient model features and leave more details, equations, and priors for App.~\ref{app:detailed-model-description}.
The primary mass distribution is a mixture of two subpopulations.
The \fsn{} subpopulation is modeled with a truncated Gaussian.  
The remainder of merging BBHs are ``distinct," and therefore we are agnostic about their origin, though we do assume they form a single population. 
For this subpopulation we consider the \bpl{} (2 power laws) and \plp{} (1 power law and 1 Gaussian) models~\cite{LIGOScientific:2025pvj}.

Schematically the mass and spin model is:
\begin{align}
    p(m_1, q, & \chieff |\lambda^f,\Lambda_{\rm FSN}, \eta_{\rm FSN}, \Lambda_{\rm X}, \eta_{\rm X} ) \nonumber\\
    &= \lambda^f \mathcal{N}\big(m_1|\Lambda_{\rm FSN} \big)\pi_0(q, \chieff | m_1, \eta_{\rm FSN}) \nonumber\\
    &+ (1 - \lambda^f) \Phi_{\rm X}(m_1|\Lambda_{\rm X})\pi_0(q, \chieff | m_1, \eta_{\rm X})\,,
    \label{eq:pop-model}
\end{align}
where $m_1$ is the primary mass, $q$ is the mass ratio, and $\chieff$ is the effective aligned spin~\cite{Damour:2001tu}, ``FSN'' denotes the \fsn{} subpopulation, and $\rm X$ stands for either ``BPL'' or ``PLP'' for the \bpl{} and \plp{} models respectively.
Among the hyperparameters, $\lambda^f$ is the fraction of sources in the \fsn{} population.
The \fsn{} primary mass distribution is a Gaussian distribution centered at $\mu^f$ with a standard deviation of $\sigma^f$, and truncated at upper and lower masses of $m^f_{\min}$ and $m^f_{\max}$, respectively. 
All other events belong to the subpopulation $\Phi_{\rm X}$, whose hyperparameters $\Lambda_{\rm X}$ are the typical parameters of the corresponding models~\cite{Talbot:2018cva, GWTC2pops, GWTC3pops, LIGOScientific:2025pvj}.

Extending previous work~\cite{Galaudage:2024meo,LIGOScientific:2025pvj}, we allow the minimum mass of the high-mass subpopulation $m_{\min}^u$ to be distinct from the minimum mass of the peak component $m^f_{\min}$. 
Specifically, the $m_{\min}^u$ prior range is $(3.5-20)\, M_{\odot}$.  
We further apply a smooth tapering function to each subpopulation.  
For the high-mass population, this means that between $m^u_{\min}$ and $m^u_{\rm turn }\equiv m^u_{\min} + \delta m^u$ the rate is suppressed by a Planck taper with scale $\delta m^u$~\cite{gwpopulation}.  
We interpret $m^u_{\rm turn}$ as the ``start'' of the high-mass subpopulation and call it the \emph{turnover mass}, one of the key targets in this work compared to, e.g., Ref.~\cite{Galaudage:2024meo}.\footnote{Our model is also qualitatively similar to the  ``2 power laws + 1 peak" model from Ref.~\cite{LIGOScientific:2025pvj}, though we treat the ``peak" as having its own spin and mass-ratio distributions, similar to Ref.~\cite{Godfrey:2023oxb}.  Both Refs.~\cite{Godfrey:2023oxb,LIGOScientific:2025pvj}, however, require the upper population to extend below the mean of the Gaussian peak, whereas we do not make this assumption.} 

The distribution $\pi_0$ represents the model for the spin and mass ratio. 
We adopt the same functional form for both subpopulations, namely a Gaussian for $\chi_{\rm eff}$ \new{(an alternative double-Gaussian model is presented in App.~\ref{app:different-spin-distribution})} and a power law for $q$ (an alternative Gaussian model is presented in App.~\ref{app:other-mass-ratio}).
The hyperparameters $\eta_{\rm FSN}$ and $\eta_{\rm X}$ of each subpopulation are distinct, motivated by evidence that the high and low masses have different mass ratio and spin distributions~\cite{Godfrey:2023oxb}. 
By not explicitly modeling precessing spins, we implicitly adopt the original parameter-estimation prior.
We further ignore potential correlations between the mass and spin~\cite{Callister:2021fpo,Biscoveanu:2021eht} and asymmetry in $\chieff$~\cite{LIGOScientific:2025pvj}, and assume the same power law redshift distribution for both subpopulations. 
These assumptions should likely have a minor impact on the low-mass sources which are our target, as they tend to be nearby (low redshift $z$) and have poorly constrained precessing spins (though see Ref.~\cite{LIGOScientific:2025brd}). 

The mass pairing deserves attention. 
Modeling the component masses $m_1$ and $m_2$ assuming they are drawn from the same distribution and then paired with some pairing function is a reasonable assumption in the case of dynamical assembly. 
In the case of isolated formation, though, this model fails for possible populations where the primary and secondary distributions have different supports.    
As pointed out in Ref.~\cite{Tiwari:2025lit}, the choice of how to model the pairing between $m_1$ and $m_2$ does affect the inferred presence of a gap, though not the $10 M_{\odot}$ peak.  
While our main mass model is written in terms of $m_1$ and $q$, in App.~\ref{app:plp-vs-pairing} we explore a model for $m_1$ and $m_2$ directly and compare with Ref.~\cite{Adamcewicz:2024jkr}.

We implement this model in \texttt{GWPopulation} \citep{gwpopulation, gwpop_pipe} and use the nested sampler \texttt{dynesty} \citep{Speagle:2019ivv} as implemented in \texttt{bilby} \citep{Ashton:2018jfp, Romero-Shaw:2020owr} to sample the hyperparameter posterior. 
We use the 153 BBHs in GWTC-4.0 that have a false alarm rate of less than $1\,\rm yr^{-1}$ and are included in the BBH analysis of Ref.~\cite{LIGOScientific:2025pvj}.\footnote{This choice is somewhat arbitrary. Events beyond GWTC-4.0, e.g.~\cite{Mehta:2025oge}, or a different threshold~\cite{Wolfe:2025yxu} could be considered, as long as sensitivity is calculated self-consistently.  We adopt these choices due to availability of sensitivity estimates and for easier comparison to other works. For the same reasons, we do not include post-GWTC-4.0 exceptional events~\cite{LIGOScientific:2025brd,LIGOScientific:2025rid}. In general, including selected exceptional results based on their measured parameters is statistically dubious. However, the events in  Refs.~\cite{LIGOScientific:2025brd,LIGOScientific:2025rid} are not inconsistent with the population inferred from GWTC-4.0~\cite{Tong:2025xir}.} \new{To ensure convergence, we require the variance of the log-likelihood in the hierarchical Monte Carlo integral for each event to be $<1.0$.  We do not impose a threshold on the number of effective Monte Carlo samples~\cite{Farr:2019rap, Talbot:2023pex}, though we verify that including such a cut weakens evidence for a gap, but leaves our qualitative conclusions intact.  That is, the posterior is still consistent with a gap at 90\% credibility. While such cuts would guarantee that all Monte Carlo integrals required for the hierarchical likelihood evaluation have converged, they do this by artificially excluding potentially high-likelihood configurations.  The only way to avoid this would be to obtain more per-event posterior and effective samples. Generally, if there were a sharp bump in the population (and then a rapid drop), it would be disfavored by such a cut. Therefore we are liberal with respect to this choice, because an alternative is not clearly available.}

Selection effects are estimated via Monte Carlo importance sampling~\cite{Tiwari:2017ndi, Farr:2019rap} based on simulated sources from an injection set~\cite{Essick:2025zed,GWTC4SensitivitySamps}.

\section{The inferred distribution of binary black hole masses}

\label{sec:failed-sn-results}

Here we present results from the \fsnplp{} (abbreviated as ``FSN+PLP'' in the figures) and the \fsnbpl{} (abbreviated as ``FSN+BPL'') models.
Both models return similar results, including consistency with a \emph{gap}, meaning an interval with strongly suppressed merger rate (quantified below), between the low-mass and the high-mass subpopulations.  
In terms of model comparison, the Bayes Factor between \fsnplp{} and \fsnbpl{} is \plpoverbplbayesfactor.
This value is neither strong nor easy to interpret, as the two models have different parameter spaces. 
The maximum likelihood ratio is \plpoverbpllikelihoodratio{}.\footnote{In App.~\ref{app:plp-vs-pairing} we examine the highest likelihood samples and find them to consistently favor the presence of a gap.}

\begin{figure*}
    \centering
    \includegraphics[width=0.99\textwidth]{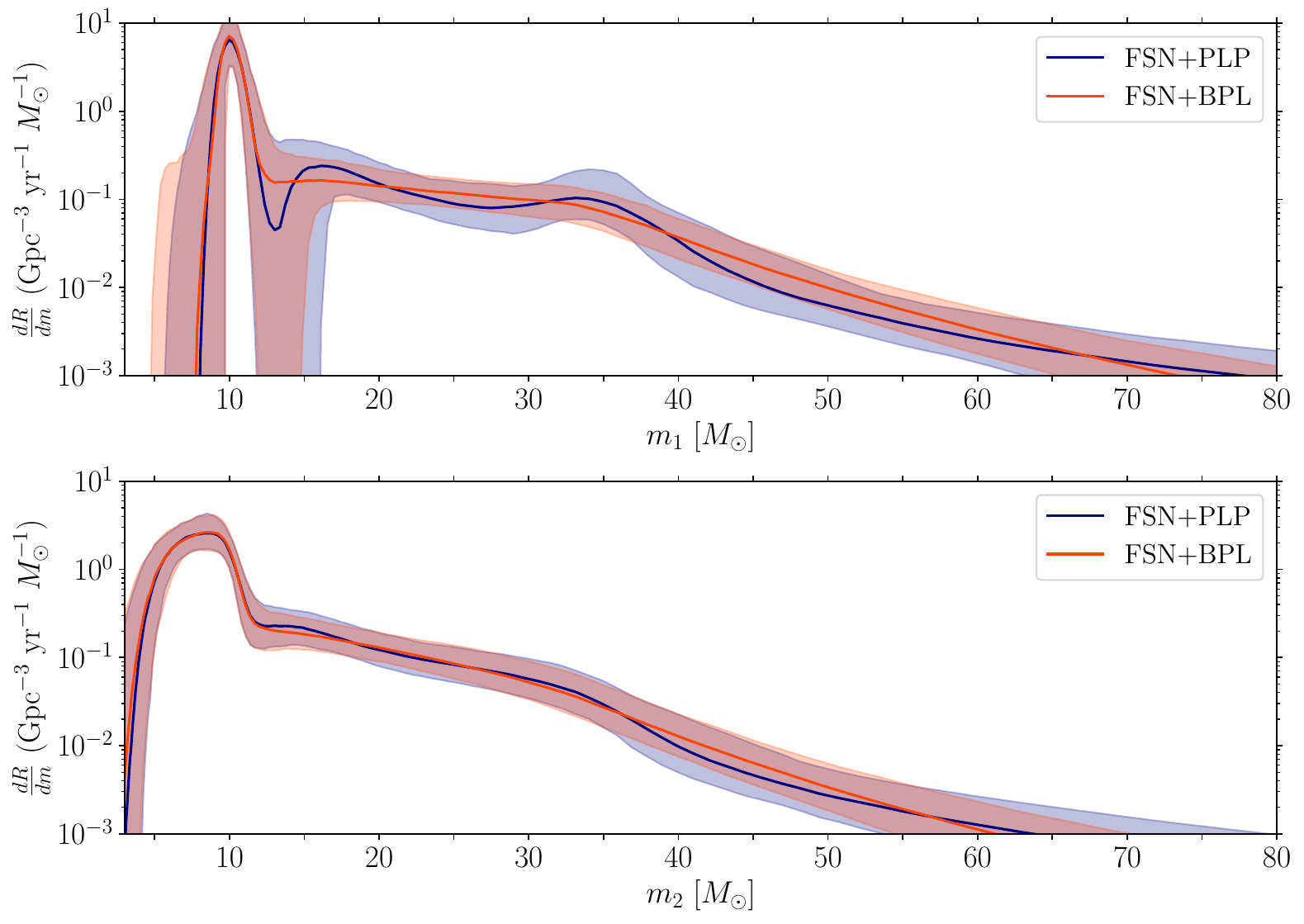}
    \caption{Posterior predictive distribution for the differential rate for the primary (top) and secondary (bottom) binary component. Results for the \fsnbpl{} model are shown in red-orange and results for \fsnplp{} are shown in blue. The 90\% symmetric credible region is shaded, while the median is marked with the corresponding color. }
    \label{fig:compare-bpl-plp-fsn}
\end{figure*}

We display the inferred differential merger rate as a function of primary and secondary mass in Fig.~\ref{fig:compare-bpl-plp-fsn} as a posterior predictive distribution (PPD).  
The PPD encodes the median and boundaries of the $90\%$ credible region of the differential rate per unit mass found by marginalizing over the model hyperparameters.
The secondary mass distribution is nearly identical between the two models and does not show evidence for a gap; we therefore focus on the primary mass which is explicitly modeled per Eq.~\eqref{eq:pop-model}.

The low-mass region around the global peak is recovered nearly identically by both models.  
In what follows, we quote numbers for the \fsnplp{} model and add the equivalent \fsnbpl{} results in parentheses.
The mean of the ${\sim} 10\, M_{\odot}$ Gaussian peak is \plpfsnbumpcenter{}  (\bplfsnbumpcenter{}) while the standard deviation is \plpfsnbumpwidth{} (\bplfsnbumpwidth{}).    
Between roughly $10$ and $14\,M_{\odot}$, the rate falls by $\sim 2$ orders of magnitude.  
Specifically, the differential rate peaks at \plpfsnbumpmaxrate{} \rateunits (\bplfsnbumpmaxrate{} \rateunits) at $\sim10\, M_{\odot}$, whereas in the range of $(13,15)\,M_{\odot}$, the highest rate is \plpmaxrateindip{} \rateunits   (\bplmaxrateindip{} \rateunits).  
The fraction of sources in the \fsn{} channel is \plpfracsourcesinpeak{} (\bplfracsourcesinpeak{}), in both cases corresponding to the majority of BBHs.
Overall, the details of the two high-mass models do not qualitatively impact the low-mass inference, especially the presence of a prominent peak and a subsequent rate drop.  
However, the choice of high-mass model does impact inference for higher masses, see App.~\ref{app:default-comparison}.

\begin{figure}
    \centering
    \includegraphics[width=0.99\linewidth]{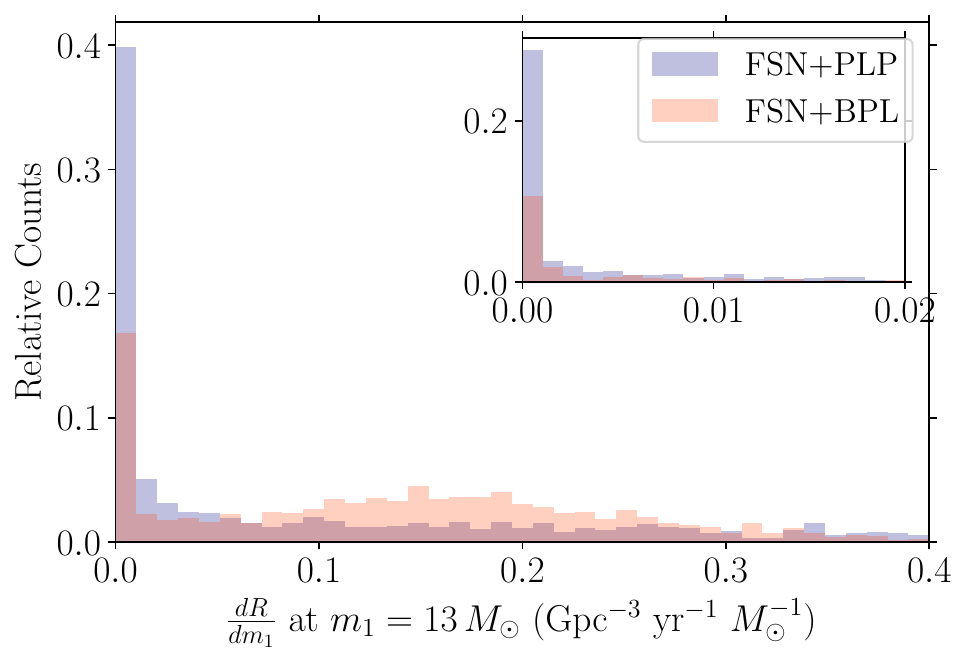}
    \caption{The inferred differential rate at $m_1=13\,M_{\odot }$ (effectively a vertical cross-section of Fig.~\ref{fig:compare-bpl-plp-fsn}) for the \fsnbpl{} and \fsnplp\, models in pink and blue respectively.  We class all samples below $10^{-3}$ \rateunits{} into a single histogram bin.  The inset zooms in near the $y$-axis and vanishing rate.}
    \label{fig:rate-at-thirteen-msun}
\end{figure}

We display the inferred differential rate at $m_1=13\,M_{\odot}$ in Fig.~\ref{fig:rate-at-thirteen-msun}.   
We define a differential rate less than $10^{-3}$ \rateunits{} as effectively zero (see below) and group such samples into a single histogram bin.  
The distributions of Fig.~\ref{fig:rate-at-thirteen-msun} have roughly two components: a ``low rate" component concentrated near zero rate and a long almost-uniform tail. 
Even the tail of the rate for this mass reaches $\sim 0.3$ \rateunits{} which is at least one order of magnitude smaller than the rate at $10\,M_{\odot}$. 
  
The question of whether a ``gap" exists in the data, in the sense of an \emph{exactly} zero differential merger rate, is impossible to answer without an infinite amount of observation time. 
Instead, we consider a criterion for the rate to be \emph{effectively zero} with current detectors and for this mass range.  
Using results from Ref.~\cite{LIGOScientific:2025slb}, we make a liberal estimate of the total amount volume-time probed for $\sim 15 + 15\,M_{\odot}$ BBHs of $\sim 20\, \rm{Gpc^3}\, \rm{yr}$ for observing runs O1--O4a, see App.~\ref{app:zero-rate-justification}.  
At a differential rate of $10^{-3} M_{\odot}^{-1} \rm{Gpc}^{-3} \rm{yr}^{-1} $, we thus would expect $\sim 0.1$ detections from an interval of $5\, M_{\odot}$ around this mass value.   
Therefore, for $\sim 12-16\,M_{\odot}$, we consider a differential rate less than $10^{-3}$ \rateunits{} to be effectively zero.\footnote{Compared to the rate at $m_1\sim 10\, M_{\odot}$, this threshold represents a drop in differential merger rate of about $10^4$.\label{footnote:rate}} 
We check that even an order of magnitude lower or higher rate threshold does not substantially alter our conclusions.  
For example, in Fig.~\ref{fig:rate-at-thirteen-msun}, grouping all sources with rate $\leq 10^{-2}$ \rateunits{} would lead to qualitatively the same picture.

With this criterion, we examine where the rate is consistent with being $\lesssim 10^{-3}$\,\rateunits{} at $90\%$ credibility. 
For the \fsnplp{} model, the rate is consistent with ``effectively zero"  in the range $m_1\in$ \plplowraterange, while for the \fsnbpl{} model the corresponding range is \bpllowraterange{}.\footnote{\new{If all Monte Carlo integrals are required to have as many effective samples as the number of included events, then under the \fsnplp{} model the corresponding range is \montecarlolimitlowraterange{}.  Therefore, the impact of such cuts based on effective samples is comparable to differences due to  model choices.} }
The data are consistent with a true ``gap" in these ranges, but do not require it, as evident from the long tail of the rate posterior in Fig.~\ref{fig:rate-at-thirteen-msun}.

\begin{table*}[]
    \centering
    \def\arraystretch{1.2}
    \begin{tabular}{|c|c|c|c|c|}
        \hline
        Model &  $m^u_{\rm turn}$ & $m^f_{\max, \rm eff}$ & $m^u_{\rm turn} -m^f_{\max, \rm eff}$ & $p(m^u_{\rm turn}>m^f_{\max, \rm eff})$   \\
        \hline
        \fsnplp{} & \plpuppermassminstar & \plpfailedmassmaxstar &  \plpdeltamaxminstar &
        \plpfracturnonabovemaxfailed{}\\
        \hline

        \fsnbpl{} &   \bpluppermassminstar & \bplfailedmassmaxstar & \bpldeltamaxminstar &
        \bplprobmturnlessmmax\\
        \hline
    \end{tabular}
    \caption{Median and $90\%$ symmetric credible region boundaries for  certain key masses for the \fsnplp{} and \fsnbpl{} models. Here $m^u_{\rm turn}$ is the turnover mass for the high-mass population and $m^f_{\max, \rm{eff}} $ is the effective maximum mass of the \fsn{} population. The latter is defined as the minimum of 3 $\sigma^f$ above the \fsn{} mean, $\mu^f$, and the \fsn{} maximum mass, $m_{\max}^f$.  We additionally display the difference between these two quantities and the probability that the turnover mass is greater than the effective maximum mass.}
    \label{tab:mass-gap-results}
\end{table*}

\begin{figure}
    \centering
    \includegraphics[width=0.99\linewidth]{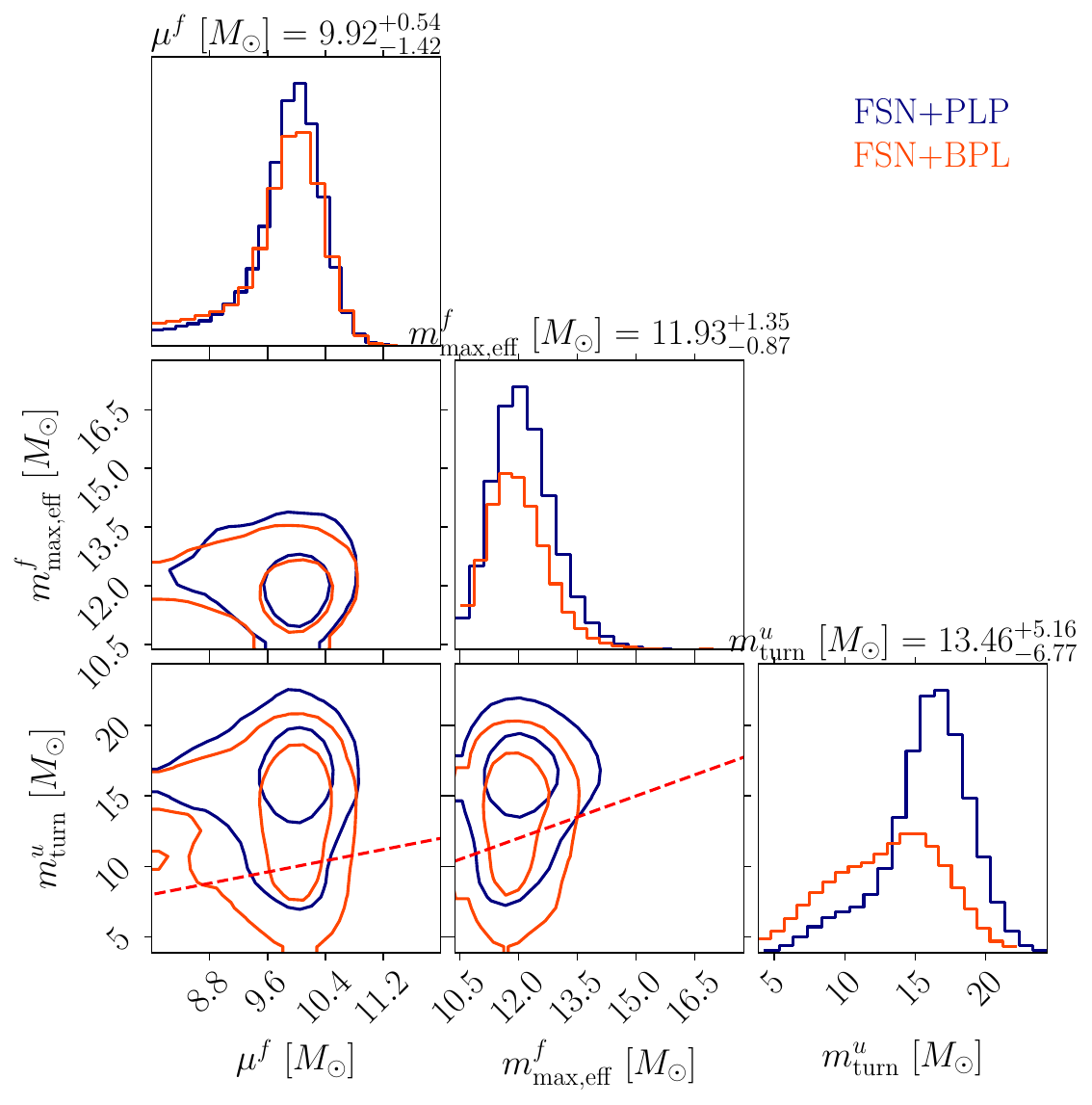}
    \caption{Marginalized posteriors on key masses related to the separations of the low- and high-mass components for the \fsnplp{} (blue) model and \fsnbpl{} (red-orange) models.  We show $\mu^f$, the mean of the \fsn{} peak, $m^f_{\max, \rm eff}$, the effective \fsn{} maximum mass (see text), and $m^u_{\rm turn}$, the turnover mass for the high-mass component. The red dashed lines mark the ``$x=y$" lines where the two quantities on each axis are equal.  Contours mark $50\%$ and $90\%$ highest-probability credible regions.}
    \label{fig:fsn-plp-key-masses}
\end{figure}

Another way to view the ``gap" is by examining whether the largest mass in the low-mass population is greater than the smallest mass in the high-mass population.
We define these two masses as follows.
Since the \fsn{} component is modeled as a truncated Gaussian, we define the \emph{effective maximum mass} of the population as $m^f_{\max, \rm eff} \equiv \min(\mu^f + 3\sigma^f, m^f_{\max})$.
This definition amounts to the largest mass in the population being either 3-$\sigma$ above the peak of the Gaussian bump or the truncation mass itself, whichever is smaller.  
As already discussed, we characterize the turnover of the high-mass population via $m^u_{\rm turn}\equiv m^u_{\min} + \delta m^{u}$, which combines the population lower truncation and the tapering scale.  
If $m^{u}_{\rm turn} > m^f_{\max, \rm eff}$, then the differential rate likely has a local minimum or ``dip" between the two masses,\footnote{Though if $m^f_{\max, \rm eff}$ and $m^u_{\rm turn}$ are close relative to the size of the taper, then there need not be a local minimum.} and if the ``dip" is sufficiently large, it will become a gap with effectively zero merger rate.  
If $m^u_{\rm turn} > \mu^f$, then there will be a distinction between the bulk of the sources in the high-mass population, and the bulk of the sources in the \fsn{} peak, even in the absence of a dip.  
In particular, $m^u_{\rm turn} > \mu^f$ would indicate that the high-mass power law does not extend to the minimum mass of the whole population.

We show the posteriors of these key masses in Fig.~\ref{fig:fsn-plp-key-masses} and quote their credible intervals in Table~\ref{tab:mass-gap-results}.  
Figure~\ref{fig:fsn-plp-key-masses} shows that for the \fsnplp{} model, the bulk of the posterior has $m^{u}_{\rm turn}$ greater than either $\mu^{f}$ or $m_{\max,\rm eff }^{f}$.
However, the posterior has a tail that renders it possible at $90\%$ credibility that the turnover mass is smaller than $\mu^f$ (which also means smaller than $m_{\max, \rm eff}^f$).  
Table~\ref{tab:mass-gap-results} quantifies these statements: $m^u_{\rm turn} -m^f_{\max, \rm eff}=$ \plpdeltamaxminstar, with $p(m^u_{\rm turn}>m^f_{\max, \rm eff})=$ \plpprobmturnlessmmax.  
The situation is qualitatively similar, albeit with less preference for a gap, for the \fsnbpl{} model as seen by the longer tail below the $m^u_{\rm turn}=m^f_{\max, \rm eff}$ line in Fig.~\ref{fig:fsn-plp-key-masses}.
Now slightly more than half (\bplprobmturnlessmmax) of the posterior supports $m^u_{\rm turn}>m^f_{\max, \rm eff}$ and $\mu^f  = m^u_{\rm turn}$ is not excluded at $50\%$ credibility. 
Nonetheless, both models prefer the turnover mass of the high-mass population to be higher than the effective maximum mass of the low-mass population, $p(m^u_{\rm turn}>m_{\max,\rm eff}^{f})\gtrsim60\%$. 
These results are consistent with the merger rate dip in Fig.~\ref{fig:compare-bpl-plp-fsn}.  
Indeed, in the \fsnplp{} model there is $\approx 88\%$ support for a local minimum in $p(m_1)$ between $10$ and $20\,M_{\odot}$ ($54\%$ for \fsnbpl{}).
Overall, a gap is supported by the majority of the posterior but not required at the 10\% credible level.

\begin{figure*}
    \centering
    \includegraphics[width=0.49\linewidth]{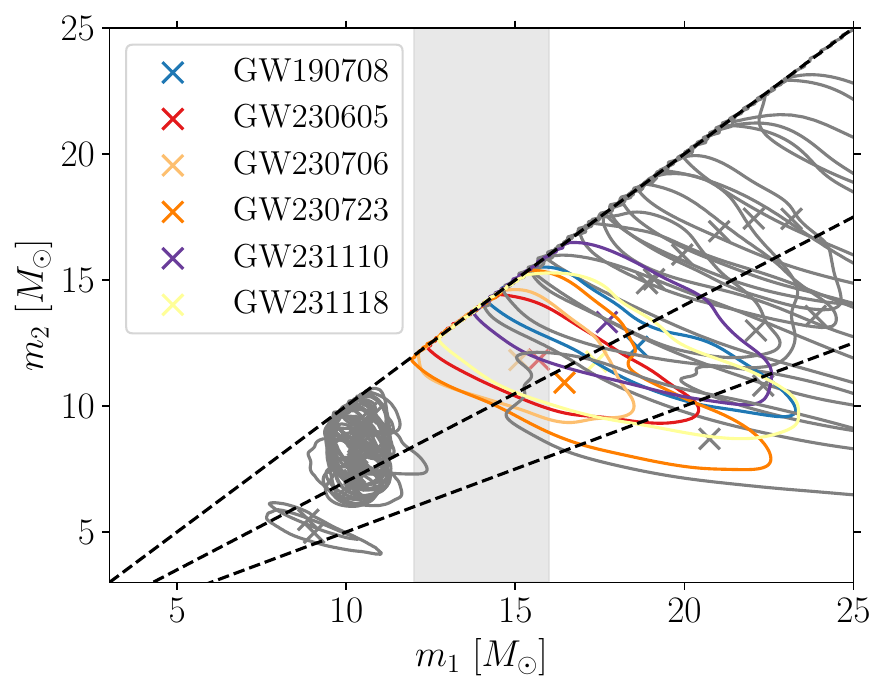}
    \includegraphics[width=0.49\linewidth]{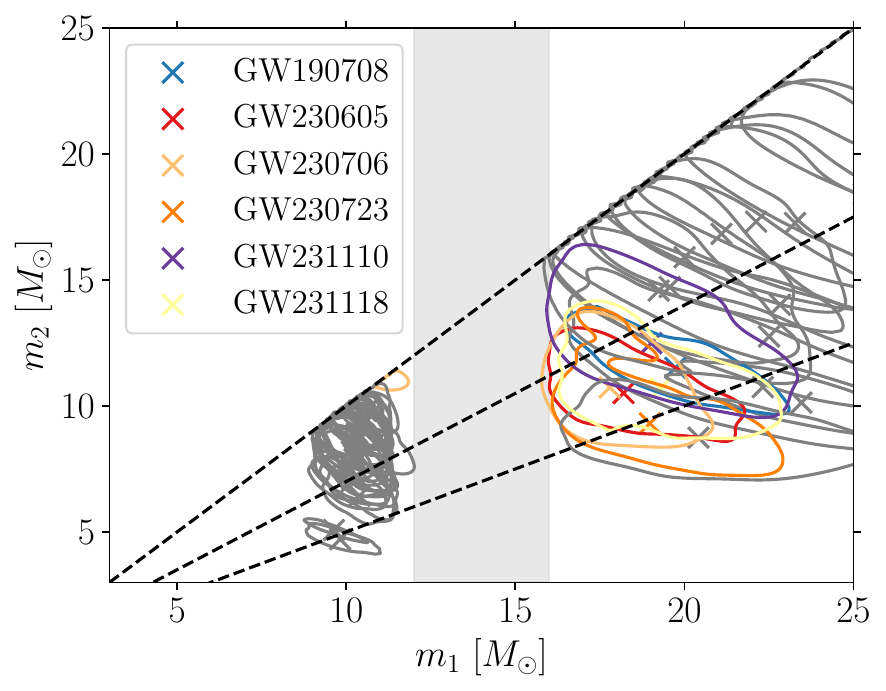}
    \caption{Marginalized posterior for the source-frame primary and secondary masses for the 153 BBHs used in this analysis under two different choices for a population prior.
    \emph{Left:} the population prior is not consistent with a gap, $m^u_{\rm turn} < 10\, M_{\odot}$.  \emph{Right:} The population prior is consistent with a gap, $m^u_{\rm turn} > 15\,M_{\odot}$.
    Contours show the 90\% posterior credible regions and stars correspond to the median.
    The shaded band denotes $(12, 16)\, M_{\odot}$, the approximate gap region under the \fsnplp{} model.
    Dashed lines denote constant mass ratio contours, $q=1.0, 0.7,$ and $0.5$.
    Most events are depicted in gray, with colors highlighting the $6$ events with the largest posterior support inside the gap.  These events are, in order, GW190708\_232457~\cite{LIGOScientific:2020ibl, LIGOScientific:2021usb}, GW230605\_065343, GW230706\_104333, GW230723\_101834, GW231110\_040320, GW231118\_090602~\cite{LIGOScientific:2025slb}. 
    Two BBHs with low masses (bottom left of each panel) do not appear by eye to belong to the $10\,M_{\odot}$ subpopulation; we discuss them in more detail in App.~\ref{app:low-mass-sources}.
    }
    \label{fig:pop-informed-pe-nogap-vs-gap}
\end{figure*}

We examine which BBHs contribute to the gap inference in Fig.~\ref{fig:pop-informed-pe-nogap-vs-gap}, where we show population-informed posteriors for the component masses of each event under two choices of population prior~\cite{Miller:2020zox,Callister:2021,Moore:2021xhn}.
This effectively replaces the default uniform parameter-estimation prior with the inferred population model. 
On the left, we select a population model (a specific sample from the population hyperposterior) that displays no gap, $m^u_{\rm turn} < \mu^f < m_{\max, \rm eff}^f$. 
On the right, we select a model with a gap, $m^u_{\rm turn} > 15\,M_{\odot} > m_{\max, \rm eff}^f$. 
Even in the no-gap case on the left panel, there is no BBH whose posterior is entirely contained in the range $m_1 \in( 12, 16)\, M_{\odot}$ at $>90\%$.
Instead there are $\sim 2$ events whose median is close to the upper edge of the gap and $\sim 4-5$ events for which $\sim 50\%$ of the posterior extends into the gap. 
A further $\sim 10$ events have lower levels of support for $m_1 \in( 12, 16)\, M_{\odot}$.   
These events are effectively ``pushed out'' of the gap on the right panel and there is reduced posterior support for $m_1 \in (12, 16)\,M_{\odot}$.  
Now, all median values lie outside of the gap region and the posteriors for $\sim 7-10$ events have low levels of support for $m_1<16\,M_{\odot}$.
This comparison provides a visual estimate for why the gap is preferred (there is no event that lies within it), but not required (a handful of events have different levels of posterior support within it).

\begin{figure*}
    \centering
    \includegraphics[width=0.49\textwidth]{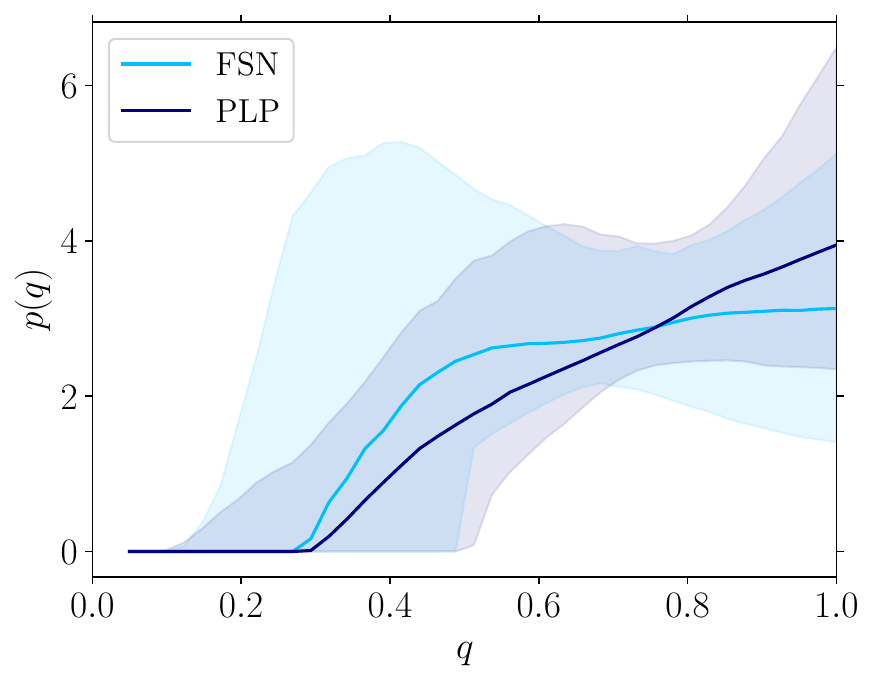}
    \includegraphics[width=0.49\textwidth]{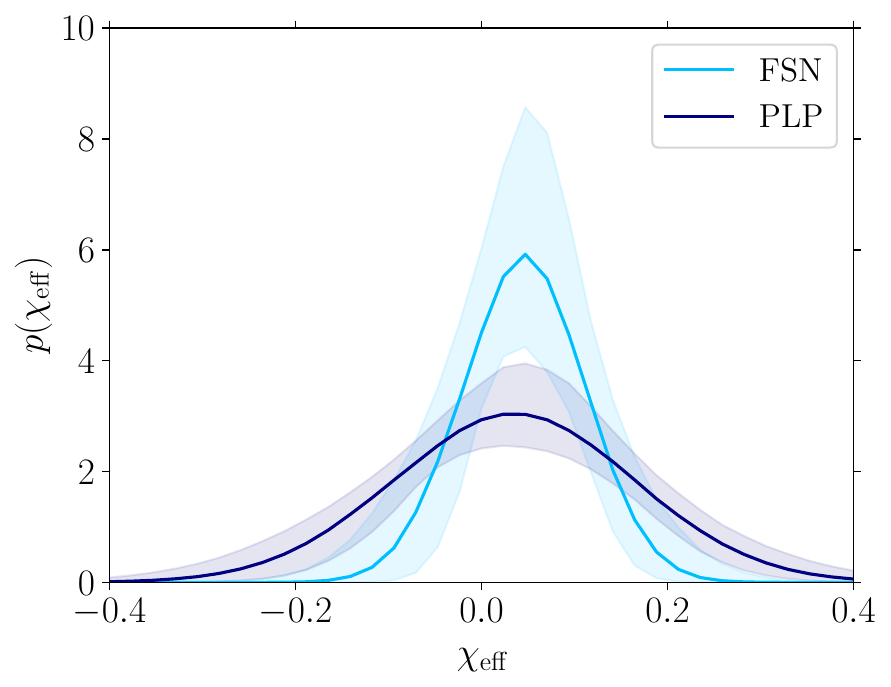}
    \caption{Inferred distribution for the binary the mass ratio $q$ (left) and the effective inspiral spin $\chieff$ (right) for the \fsn{} and \plp{} subpopulations of the \fsnplp{} model in light and dark blue respectively.}
    \label{fig:fsn-plp-q-chieff}
\end{figure*}

Finally, we contrast the inferred mass ratio and spin properties of the low- and high-mass subpopulations.
In Fig.~\ref{fig:fsn-plp-q-chieff} we show results for the \fsnplp{} model, but find a qualitatively consistent picture with the \fsnbpl{} model. 
This is because across the two models, each subpopulation contains effectively the same sources.  

Starting with the mass ratio distributions, the two subpopulations are similar within uncertainties, also consistent with Ref.~\cite{LIGOScientific:2025pvj}.
The power law slope is $\beta^f =$ \plpfailedbeta{} and $\beta^u=$ \plpupperbeta{} for the low- and high-mass subpopulations respectively.  
The \fsn{} subpopulation tends to have more asymmetric binaries, $\beta^u > \beta^f$ at \result{$73\%$} credibility.   
The two subpopulations also have nearly identical cutoffs, $q_{\min}^f =$ \plpfailedqmin{} and $q_{\min}^u = $ \plpupperqmin{}.  
Therefore, on average, the \fsn{} subpopulation tends to have a broader mass ratio distribution and a sharper truncation.\footnote{These results are likely affected by the fact that binaries with secondaries which cannot be unambiguously distinguished from neutron stars have been excluded from the dataset we analyze. Therefore the inference of $q^f_{\min}$, and potentially $\beta^f$, should be interpreted with caution.}  
In App.~\ref{app:other-mass-ratio} we explore a Gaussian model for the mass ratio of the \fsn{} subpopulation.
This is motivated by the consideration that since the primary mass has a preferred value, then that may also be the case for the secondary.  
Furthermore, the \fsn{} mass ratio distribution depends strongly on a pair of sources that are likely outliers, see Fig.~\ref{fig:pop-informed-pe-nogap-vs-gap}; we investigate them further in App.~\ref{app:low-mass-sources}.

Switching to the spin, we find substantially different $\chieff$ distributions between the \fsn{} and the high-mass subpopulations, consistent with Refs.~\cite{Godfrey:2023oxb, Banagiri:2025dmy}.  
The \fsn{} subpopulation has a narrower $\chieff$ distribution peaked at \fsnplpbumpchieffmean{} and with a standard deviation of \fsnplpbumpchieffsigma{}.
The high mass distribution peaks at a similar value, \fsnplpupperchieffmean{}, but it is wider with a standard deviation of \fsnplpupperchieffsigma{}.

\section{Discussion and Comparison to Past Work}
\label{sec:conclusions}

We have proposed and explored an extension to population models for the BBH mass distribution inspired by SNe simulations. 
We considered the possibility that failed SNe are responsible for the observed abundance of GW sources with primary masses $\sim 10\,M_{\odot}$, a channel that remains inoperative at adjacent masses.  
The observed drop in BBH merger rate by 1-2 orders of magnitude within a few solar masses is then consistent with results from \emph{ab initio} SNe simulations of solar-metallicity progenitors where a narrow ZAMS mass range that fails to explode maps onto the observed BH peak. 

Analyzing the current catalog of BBH detections, we find that the rate of mergers with primary masses $\sim12-16\,M_{\odot}$ is consistent with zero, however such a ``gap'' is not yet unambiguously preferred by the data.
Specifically, the higher-mass population truncates at \plpuppermassminstar{}, which at $93\%$ credibility is larger than the mean primary mass in the peak, \plpfsnbumpcenter{}.  
Moreover, there is a $\result{88\%}$ probability for a separation or ``gap" between the $10\, M_{\odot}$ peak and the rest of the high-mass sources, Table~\ref{tab:mass-gap-results}.
If such a gap is confirmed with future data, it would constitute strong evidence that the $\sim 10\,M_{\odot}$ BBHs originate from an astrophysically distinct mechanism than the remaining high-mass population.

The presence of substructure in the BBH primary mass distribution beyond a power law has been established in numerous other studies.
\citet{Farah:2023vsc, Godfrey:2023oxb} both found that the peak at $\sim 10 M_\odot$, first identified in Refs.~\cite{Edelman:2021zkw, Edelman:2022ydv}, is robust under statistical noise or modeling systematics.
These studies further noted that the peak may be more than the global maximum from a power law continuum, meaning that the peak is a genuine feature and not the onset of a global power law.

Multiple studies have further shown suppression of the merger rate in $\sim12-16\,M_{\odot}$, though to varying degrees.  
Using data from GWTC-3.0~\cite{KAGRA:2021vkt,GWTC3pops}, and a nonparametric model based on splines, \citet{Edelman:2022ydv} showed that the merger rate is suppressed by a factor of $\sim 100$ between the $10\,M_{\odot}$ peak and a trough near $\sim 15\, M_{\odot}$.  
Using the same data and a variational autoencoder model, \citet{Callister:2023tgi} also found a rate suppression, but of a factor of $\sim 10$, and argued that a single power law in this mass range would overpredict the merger rate in the $\sim 15-30\,M_\odot$ range by a factor of $\sim 2-3$.   
These studies, and further nonparametric analyses~\cite{Edelman:2021zkw,Farah:2023vsc}, do not find evidence for a gap in the sense of zero merger rate.  
\citet{Bertheas:2026odj} found indications of a strongly suppressed merger rate between $12\,M_{\odot}$ and $16\,M_{\odot}$ using a flexible population composed of multiple power laws with GWTC-4.0 data, but they did not explicitly model spins as their primary objective was inferring cosmological parameters.
However, the mass distribution features under question involve very sharp rate changes.
Nonparametric models would have to \emph{a priori} allow for changes of the merger rate by $\sim 10^4$ over the range of a few solar masses to identify a gap.\textsuperscript{\ref{footnote:rate}} 

The primary mass clustering around $10\,M_{\odot}$ is not the only unique characteristic of the peak-residing BBHs.
Based on GWTC-3.0 data and a splines mass model, \citet{Godfrey:2023oxb} showed that BBHs in the $10\,M_{\odot}$ peak have a different spin distribution than those at higher masses and argued for an isolated formation origin.  
They did not explicitly search for a gap.
Based on the same data, \citet{Li:2022jge} also identified a low-mass subpopulation with a unique mass ratio distribution, and spin distribution~\cite{Li:2024jzi}.   
These results were confirmed by \citet{Banagiri:2025dmy} who, using GWTC-4.0 data, identified a ``low-mass population" with unique mass ratio and spin characteristics extending up to $\sim 28\,M_{\odot}$, also confirmed in \citet{Ray:2026uur}. 
They also did not explicitly search for a gap.  
Our results confirm that the $10\,M_{\odot}$ binaries have a narrower aligned spin distribution than the rest of the population. 
Since our result is consistent with a gap, however, it is possible the two subpopulations may be distinct in both origin and primary mass range.   

\citet{Galaudage:2024meo} connected the $\sim 10 \,M_{\odot}$ subpopulation with BHs formed from stars with high progenitor core compactness~\cite{Schneider:2023mxe, Willcox:2025cus,Willcox:2025poh}.
Based on GWTC-3.0 data, they also identified a suppressed merger rate of order $\sim 100$ from the peak of the $10\, M_{\odot}$ to the trough at $\sim 14-15\, M_{\odot}$. 
This suppression is limited in their model by the presence of a global power law which extends across the full mass range, precluding a gap.
This modeling choice is driven by the aim to look for both a core-compactness bump due to carbon-oxygen burning, and an additional Gaussian bump due to oxygen-neon burning at higher mass.  
As discussed in Sec.~\ref{sec:introduction}, however, it is unlikely that core compactness that predicts a rate enhancement of $\sim2-3\times$~\cite{Schneider:2023mxe, Willcox:2025cus,Willcox:2025poh} alone can explain the observed enhancement of $\gtrsim50\times$ the rate at $\sim 10\, M_{\odot}$.  

\citet{Tong:2025xir} explored the potential of higher-generation mergers in the relevant mass region based on GWTC-4.0 data and the special events GW241011 and GW241110~\cite{LIGOScientific:2025brd} (which we do not use).
The mass model was a power law plus two peaks (and a high-mass gap in $m_2$~\cite{Tong:2025wpz}) and higher-generation mergers were targeted via the spin model (Gaussian vs uniform).
They identified a suppressed first-generation merger rate in the range $\sim(12, 20)\, M_{\odot}$, which they speculate may be due to differences in core compactness of progenitors, i.e., the same mechanism as Ref.~\cite{Galaudage:2024meo}.
No gap in the total rate was identified nor explicitly modeled.  
Further, \citet{Li:2025iux}, \citet{Farah:2026jlc} and \citet{Ray:2026uur} also identified a potential hierarchical subcomponent of the population peaking at $\sim 16\,M_{\odot}$.  
Therefore, explicitly modeling mergers $m_1\sim 16\,M_{\odot}$ via their spin may further increase evidence for a gap in the distribution of first-generation mergers.  
  
To our knowledge, the only parametric study that explicitly looked for a gap in the $12-15\,M_{\odot}$ range is \citet{Adamcewicz:2024jkr} using GWTC-3.0 data.
For this study, the standard mass model of a power law with two peaks was augmented by a notch filter whose amplitude controls the magnitude of the merger rate dip.
Unlike our differential merger rate in Fig.~\ref{fig:compare-bpl-plp-fsn} that is consistent with a gap, they find a nonzero merger rate throughout.
We attribute this discrepancy to modeling choices: while we look for a gap in the primary mass,  \citet{Adamcewicz:2024jkr} imposed the same underlying mass distribution for $m_1$ and $m_2$ before weighing by a pairing function $f(m_2/m_1)$.
Such a model does not permit the rate of mergers to be zero in, e.g., $m_1$ without it also being zero in $m_2$, since $p(m_1)\propto p(m_2)f(m_2/m_1)$.  
We verify that we can reproduce these results in App.~\ref{app:plp-vs-pairing} by using a qualitatively similar mass model.
Nonetheless, \citet{Adamcewicz:2024jkr} do find a rate suppression by a factor of $\sim 20-30$ between $\sim 10\,M_{\odot}$ and $\sim 15\, M_{\odot}$. 
Therefore the main difference between our result and theirs is the long tail of the rate distribution down to effectively zero in the range $\sim 12-16\, M_{\odot}$ in Fig.~\ref{fig:compare-bpl-plp-fsn}.
This result and our App.~\ref{app:plp-vs-pairing} confirm that if there is a gap in the $m_1$ distribution, it is not coincident with a gap in the $m_2$ distribution.  

It is possible, however, that $m_1$ and $m_2$ display suppressed merger rates in different mass ranges, e.g., the recent evidence for a pair instability gap in $m_2$ only~\cite{Antonini:2025zzw,Roy:2025ktr,Tong:2025wpz,Antonini:2025ilj,Wang:2025nhf,Ray:2025xti,Sridhar:2025kvi}.
Identifying differences in the $m_1$ and $m_2$ distributions and connecting them to astrophysical models is complicated by \emph{label-swapping}~\cite{Biscoveanu:2021eht, Golomb:2024mmt, Gerosa:2024ojv}.
The GW signal, and thus the analysis likelihood, is invariant under exchanging the binary components, with the convention $m_1>m_2$ breaking this invariance.
Consider the case of binaries constructed by sampling two masses $m_A,m_B$ independently and identically from the same astrophysical distribution and then sorting them to produce a pair $m_1>m_2$.  
The random variables $m_1$ and $m_2$ do not inherit the distributions of $m_A$ and $m_B$, but rather $m_2$ ($m_1$) is preferentially smaller (larger) than the central value of $m_A$ or $m_B$. 
Therefore, the distribution of $m_1$'s may not reflect the distribution of any particular astrophysical class of BHs.  
Using total mass (or chirp mass) would reduce the temptation to make this association, but at the cost of losing the interpretability of $m_1$ as a ``typical," if perhaps biased, component mass.  
This problem is worse for poorly measured binary mass ratios and can be approached by alternative classification methods~\cite{Gerosa:2024ojv}, spin sorting~\cite{Biscoveanu:2021eht}, tides sorting~\cite{Golomb:2024mmt}, or ``doubling'' the posterior by exchanging $1\leftrightarrow 2$ if there is no motivation to distinguish the components.

Returning to our results, if the turnover of the high-mass power law is indeed higher than $10\, M_{\odot}$, a plausible astrophysical interpretation is that a very particular channel is responsible for the $\sim 10\, M_{\odot}$ BBHs~\cite{Burrows:2024wqv}\footnote{If failed SNe are the origin of such a feature, fully general relativistic simulations of BH formation~\cite{Kuroda:2023mzi} from failed SNe will likely be necessary to determine the finer details of this distribution theoretically.}.   
Given current uncertainties, a more conservative interpretation would be that the rate of mergers with primaries in $\sim 12- 16 \,M_{\odot}$ is orders of magnitude lower than the rate at $\sim 10\,M_{\odot}$, but its lower limit is poorly constrained.  
Models that correlate the rate in this region with higher masses using, e.g., a global power law that cannot terminate in this region or a nonparametric model with a large correlation length, may be systematically overestimating the rate, e.g., Ref.~\cite{Callister:2023tgi} and App.~\ref{app:default-comparison}.  
As an example, in App.~\ref{app:default-comparison} we examine the headline GWTC-4.0 population model~\cite{LIGOScientific:2025pvj}, and show how the $\sim12-16\,M_{\odot}$ modeling can impact results for higher masses. 
Model construction is not solved merely by model selection.  
General caveats about the interpretation of Bayes Factors between models with widely different parameter spaces are discussed, e.g., in Apps.~\ref{app:default-comparison} and~\ref{app:other-mass-ratio}.
Maximum-likelihood estimates suffer from their own ambiguities.
\citet{Guttman:2025jkv} computed the global \emph{optimal} population distribution that maximizes the likelihood. 
This distribution consists of a finite sum of delta functions (for a finite number of sources), thus there is no canonical way to decide if a separation between two components of the population is significant given data alone.
More targeted approaches that examine specific features of models, such as posterior predictive checks~\cite{Fishbach:2019ckx,Romero-Shaw:2022ctb,WinneyPPC}, are more suitable for scrutinizing specific features of models.    

\new{
While this study was under review, the latest GWTC-5.0 catalog of BBH detections was released~\cite{LIGOScientific:2026wfs,LIGOScientific:2026ctl}.
We confirm that our headline results hold with this more extended dataset by repeating the headline analysis of \fsnplp{} using the 257 BBHs of GWTC-5.0 that pass our false-alarm-rate threshold (and excluding two past events that were first identified in GWTC-5.0).
The component mass distribution is consistent with the GWTC-4.0 one of Fig.~\ref{fig:compare-bpl-plp-fsn}.
Uncertainties overall shrink and the prominence of the $\sim 35\,M_{\odot}$ peak decreases, as also reported in Ref.~\cite{LIGOScientific:2026ctl}.
The mass ratio and spin distributions remain consistent with those of Fig.~\ref{fig:fsn-plp-q-chieff}, and specifically the \plp{} $\chieff$ distribution is definitively wider than the \fsn{} one.
Support for the gap further persists.
The probability of $m^u_{\rm turn}>m^f_{\max, \rm eff}$ is $0.85$, adjusted from \plpprobmturnlessmmax\, in GWTC-4.0, and 90\% of the posterior displays a local minimum in $p(m_1)$.
The primary mass rate is consistent with ``effectively zero'' in the range $(12.3-15.7)M_{\odot}$.

These results are again obtained with a maximum variance of the Monte Carlo integrals of less than 1.0.
We have confirmed that they are largely unaffected by this choice, as relaxing the threshold to 4.0 leaves the mass range of ``effectively zero" rate unchanged and adjusts the probability of a local minimum in $p(m_1)$ to $88\%$.
Consistent with Fig.~15 of Ref.~\cite{LIGOScientific:2026ctl}, a weaker variance threshold also leads to a marginally smaller width for the $10\,M_{\odot}$ peak.
If Monte Carlo convergence were instead imposed based on the total number of events~\cite{Farr:2019rap}, the impact would be likely more noticeable due to the larger number of events (257) in GWTC-5.0.}


Finally, simulations also suggest that some BHs are formed from successful SNe~\cite{Chan:2017tdg, Woosley:1996uu, Moriya:2019jon, Burrows:2024wqv}.  
Such a mechanism complicates the relation between ZAMS mass, stellar mass at collapse, and remnant compact object mass, and may lead to low-mass $\sim 3\, M_{\odot}$ BHs.      
We do not model such BHs here as it is not straightforward to distinguish them from NSs~\cite{LIGOScientific:2020zkf,LIGOScientific:2024elc}.  
Therefore, the true underlying structure of the mass distribution may be more complicated.
Effort to identify gaps, dips, and in general features in the mass distribution will continue offering clues about subpopulations with potentially distinct astrophysical origins.

\section{Acknowledgments}

The authors thank Ethan Payne, Rhiannon Udall, Tianshu Wang, Hui Tong, Aditya Vijaykumar, and Amanda Farah for insightful conversations.
This work was supported by NSF within the framework of the MUSES collaboration, under grant number OAC-2103680.  
I.L. and K.C. acknowledge support from
the Department of Energy under award number DESC0023101 and by a grant from
the Simons Foundation (MP-SCMPS-00001470).
J.G. acknowledges the support from the National Science Foundation through the Grant NSF PHY-2207758.
This material is based upon work
supported by NSF’s LIGO Laboratory which is a major facility fully funded by the National Science Foundation.
The authors are grateful for computational resources provided by the LIGO Laboratory and supported by National Science Foundation Grants PHY-0757058 and PHY-0823459. 
This work made use of the {\sc numpy}~\cite{harris2020array}, {\sc scipy}~\cite{2020SciPy-NMeth}, and {\sc matplotlib}~\cite{Hunter:2007} python modules. The 
{\sc gwpopulation}~\cite{gwpopulation} mass models and {\sc gwpopulation-pipe}~\cite{gwpop_pipe} configuration files used in this work, as well as hyperposterior samples generated will be made available upon publication at the following DOI:  \href{https://doi.org/10.5281/zenodo.21248115}{10.5281/zenodo.21248115}.

\appendix

\section{Detailed description of the population model}
\label{app:detailed-model-description}

\begin{table*}[]
\centering
\caption{Hyperparameter definitions and prior distributions for the \fsn{} (first block) and \plp{} (second block) subpopulations of the \fsnplp {} model, as well as the joint parameters (third block).  The fraction of sources in the \plp{} channel is $\lambda^u = 1-\lambda^f$. }
\label{tab:prior-plp}
\begin{tabular}{lcccc}
\textbf{Parameter} & \textbf{Population} & \textbf{Range} & \textbf{Distribution} & \textbf{Description} \\
\hline
$\mu^f$        & FSN & (8, 12) $[M_{\odot}]$       & Uniform & Mean of primary mass peak \\
$\sigma^f$     & FSN & (0.25, 4) $[M_{\odot}]$    & Uniform & Standard deviation of primary mass peak \\
$m_{\max}^f$        & FSN & (12, 20) $[M_{\odot}]$      & Uniform & Maximum BH mass \\
$m_{\min}^f$        & FSN & (5, 12) $[M_{\odot}]$       & Uniform & Minimum BH mass \\
$\mu_{\chi}^f$      & FSN & (0, 1)        & Uniform & Mean of $\chieff$ distribution \\
$\sigma_{\chi}^f$   & FSN & (0, 1)        & Uniform & Standard deviation of $\chieff$ distribution \\
$\beta^f$       & FSN & (-2, 2)      & Uniform & Mass ratio power-law slope \\

$q_{\min}^f$        & FSN & (0.05, 0.9)   & Uniform & Minimum mass ratio \\
$\delta_m^f$        & FSN & (0.25, 10) $[M_{\odot}]$   & Uniform & Taper scale in mass spectrum at low mass \\
\hline
$\alpha^u$        & PLP & (-4, 12)     & Uniform & Power-law slope of primary mass \\
$\lambda_{\rm pp}^u$         & PLP & (0, 1)        & Uniform & Fraction of PLP sources in the Gaussian  \\
$\beta^u$         & PLP & (-2, 7)        & Uniform & Mass ratio power-law slope \\
$m_{\max}^u$        & PLP & (30, 170) $[M_{\odot}]$     & Uniform & Maximum BH mass \\
$m_{\min}^u$        & PLP & (3.5, 20) $[M_{\odot}]$ & Uniform & Minimum BH mass \\
$\mu_{\chi}^u$      & PLP & (0, 1)        & Uniform & Mean of spin distribution \\
$\sigma_{\chi}^u$   & PLP & (0, 1)        & Uniform & Standard deviation of spin distribution \\
$\mu^u_{\rm pp}$        & PLP & (20, 50) $[M_{\odot}]$      & Uniform & Mean of the Gaussian in mass \\
$\sigma^u_{\rm pp }$     & PLP & (0.25, 20) $[M_{\odot}]$    & Uniform & Standard deviation of the Gaussian in mass \\
$q_{\min}^u$        & PLP & (0.05, 0.9)   & Uniform & Minimum mass ratio \\
$\delta_m^u$        & PLP & (0.25, 10) $[M_{\odot}]$   & Uniform & Taper scale in mass spectrum at high masses  \\
\hline
$\lambda^f$               & FSN & (0, 1)        & Uniform & Fraction of sources in the FSN channel \\
$\lambda_z$       & --  & (-10, 10)     & Uniform & Redshift evolution parameter \\
\hline
\end{tabular}
\end{table*}

\begin{table*}[]
\centering
\caption{Hyperparameter definitions and prior distributions for the \fsn{} (first block) and \bpl{} (second block) subpopulations of the \fsnbpl{} model, as well as the joint parameters (third block).  The fraction of sources in the \bpl{} channel is $\lambda^u = 1-\lambda^f$.}
\label{tab:prior-bpl}
\begin{tabular}{lcccc}
\textbf{Parameter} & \textbf{Population} & \textbf{Range} & \textbf{Distribution} & \textbf{Description} \\
\hline
$\mu^{f}$          & FSN & (8, 12) $[M_{\odot}]$      & Uniform & Mean of primary mass peak \\
$\sigma^{f}$       & FSN & (0.25, 4) $[M_{\odot}]$      & Uniform & Standard deviation of primary mass peak \\
$m^{f}_{\max}$          & FSN & (12, 20)  $[M_{\odot}]$       & Uniform & Maximum BH mass \\
$m^{f}_{\min}$          & FSN & (5, 12)  $[M_{\odot}]$       & Uniform & Minimum BH mass \\
$\mu^{f}_{\chi}$        & FSN & (0, 1)         & Uniform & Mean of $\chieff$ distribution \\
$\sigma^{f}_{\chi}$     & FSN & (0, 1)         & Uniform & Standard deviation of $\chieff$ distribution \\
$\beta^{f}_{q}$         & FSN & (-2, 2)        & Uniform & Mass-ratio power-law slope \\
$q_{\min f}$            & FSN & (0.05, 0.9)    & Uniform & Minimum mass ratio \\
$\delta_{m f}$          & FSN & (0.25, 10)  $[M_{\odot}]$   & Uniform & Taper scale in mass spectrum at low mass \\
\hline
$\alpha^{u}_{1}$        & BPL & (-4, 12)       & Uniform & Power-law slope below mass break \\
$\alpha^{u}_{2}$        & BPL & (-4, 12)       & Uniform & Power-law slope above mass break \\
$\beta^{u}_{q}$         & BPL & (-2, 7)        & Uniform & Mass-ratio power-law slope \\
$m^{u}_{\max}$          & BPL & (30, 170)  $[M_{\odot}]$     & Uniform & Maximum BH mass \\
$m^{u}_{\min}$          & BPL & (3.5, 20)  $[M_{\odot}]$     & Uniform & Minimum BH mass \\
$b^{u}$                 & BPL & (0, 1)         & Uniform & Fraction of systems above mass break \\
$\mu^{u}_{\chi}$        & BPL & (0, 1)         & Uniform & Mean of spin distribution \\
$\sigma^{u}_{\chi}$     & BPL & (0, 1)         & Uniform & Standard deviation of spin distribution \\
$q^u_{\min}$            & BPL & (0.05, 0.9)    & Uniform & Minimum mass ratio \\
$\delta^u_{m}$          & BPL & (0.25, 10) $[M_{\odot}]$     & Uniform & Taper scale in mass spectrum \\
\hline
$\lambda^{f}$           & FSN & (0, 1)         & Uniform & Fraction of sources in the FSN channel \\
$\lambda_z$             & --  & (-10, 10)     & Uniform & Redshift evolution parameter \\
\hline
\end{tabular}
\end{table*}

In this Appendix we present more technical details about the population models described in Sec.~\ref{sec:failed-sn-models}. 
The mass and spin model is given in Eq.~\eqref{eq:pop-model}, which we reproduce here for completeness:
\begin{align}
    p(m_1, q, & \chieff |\lambda^f,\Lambda_{\rm FSN}, \eta_{\rm FSN}, \Lambda_{\rm X}, \eta_{\rm X} ) \nonumber\\
    &= \lambda^f \mathcal{N}\big(m_1|\Lambda_{\rm FSN} \big)\pi_0(q, \chieff | m_1, \eta_{\rm FSN}) \nonumber\\
    &+ (1 - \lambda^f) \Phi_{\rm X}(m_1|\Lambda_{\rm X})\pi_0(q, \chieff | m_1, \eta_{\rm X})\,.
    \label{eq:pop-model-detailed}
\end{align}
The spin and mass ratio distribution is given by 
\begin{equation}
    \pi_0(q, \chieff| \eta) \propto q^{\beta} \Theta(q - q_{\min}) \mathcal N(\chieff | \mu_{\chi}, \sigma_{\chi})\,,
\end{equation}
where the hyperparameters $\eta$ are allowed to be different for the \fsn{} and high-mass populations, $\eta_{\rm FSN}$ and $\eta_{\rm X}$ respectively. 
Their definition and prior are given in Tables~\ref{tab:prior-plp} and~\ref{tab:prior-bpl}.

The \plp{} model has the form 
\begin{multline}
    \label{eq:plp-form}
    \Phi_{\rm PLP}(m_1|\Lambda_{\rm PLP}) =\\(1-\lambda_{\rm pp}^u) K(\alpha^u) m_1^{\alpha^u} + \lambda_{\rm pp}^u \mathcal N(m_1| \mu_{\rm pp
    }^u, \sigma_{\rm pp}^u)\,,
\end{multline}
for $m_{\min}<m <m_{\max}$ and zero elsewhere, and $K(\alpha)$ is a normalization factor.
The \bpl{} model has the form 
\begin{multline}
    \label{eq:bpl-form}
    \Phi_{\rm BPL}(m_1|\Lambda_{\rm BPL}) =\\\begin{cases}
    K_1(\alpha_1^u, \alpha_2^u, m_{\rm b}^u)m_1^{\alpha_1^u};\ \ m_1<m_{\rm b}^u\\
    K_2(\alpha_1^u, \alpha_2^u, m_{\rm b}^u) m_1^{\alpha_2^u};\ \  m_1 > m_{\rm b}^u
    \end{cases}\,, 
\end{multline}
for $m_{\min}<m <m_{\max}$ and zero elsewhere, with $K_1$ and $K_2$ being normalization factors.  The break mass is not sampled, but computed by enforcing a certain fraction of sources, $b^u$ (which is sampled) are above the break.\footnote{See App.~B of Ref.~\cite{LIGOScientific:2025pvj} and references therein for details.} 
We apply smooth Planck tapers of width $\delta_m$ separately to all subpopulations~\cite{gwpopulation} which we also infer.
The definition and prior ranges are given for the \fsnplp{} model in Table~\ref{tab:prior-plp} and for the \fsnbpl{} model in Table~\ref{tab:prior-bpl}.

\section{Comparison with the Default model from GWTC-4.0}
\label{app:default-comparison}

\begin{figure*}
    \centering
    \includegraphics[width=0.99\textwidth]{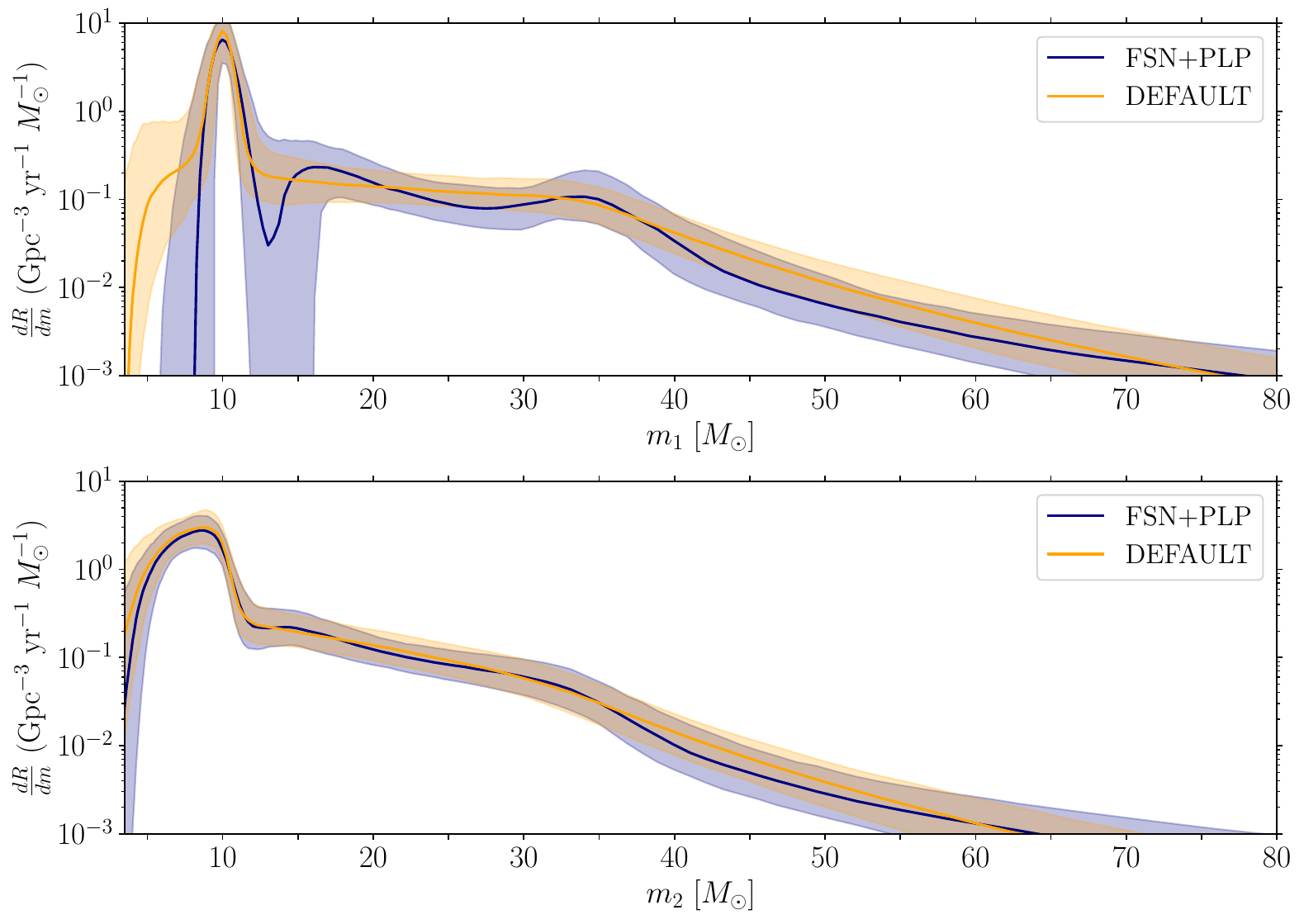}
    \caption{Same as Fig.~\ref{fig:compare-bpl-plp-fsn}, but comparing the \fsnplp{} (blue) to the \default{} (yellow) model of Ref.~\cite{LIGOScientific:2025pvj} that does not allow for a gap in the primary mass distribution.}
    \label{fig:plp-vs-default-m1m2-envelope}
\end{figure*}

In this Appendix, we compare our results to the ``2 power law + 2 peak" model from Ref.~\cite{LIGOScientific:2025pvj}, which we call the \default\, model. 
Relative to that implementation, we make some minor adjustments to increase consistency with our work, including tapering of the minimum mass and adopting our Gaussian spin distribution.  
Overall, the \default\, model is similar to the \fsnplp{}\, model with some key differences. 
First, the \default{} model uses a unified spin and mass-ratio distribution across the mass range, while the \fsnplp{} model allows for separate distributions for the \fsn{} and high-mass subpopulations. 
Second, while both models include a Gaussian peak at $\sim 10\, M_{\odot}$, at higher masses the \fsnplp{} model uses a power law + peak, while the \default{} model uses a \emph{broken} power law + peak.  
Third, and most importantly for our purposes, \fsnplp{} allows for a gap in the primary mass distribution between the $\sim 10\, M_{\odot}$ peak and higher masses, which  the \default{} model does not permit.

We show the differential merger rate as a function of component masses in Fig.~\ref{fig:plp-vs-default-m1m2-envelope}.  
The most evident differences between the two models are in the primary mass distribution: the \default{} model has no dip around $13\,M_{\odot}$ (because such a feature is not allowed) and a ``shoulder" at $m_1 \lesssim 9\,M_\odot$.  
The latter is likely due to two (potentially anomalously) low-mass BBH systems that the \fsnplp{} model includes in the \fsn{} subpopulation, see App.~\ref{app:low-mass-sources}.  

There are further more subtle differences between the two distributions.  
In particular, the \default{} model, much like \fsnbpl{} model in Fig.~\ref{fig:compare-bpl-plp-fsn}, shows little structure in the mass distribution between $\sim16\,M_{\odot}$ and $\sim35\,M_{\odot}$, being distributed as a power law.  However, above $\sim 35\,M_{\odot}$, a notably different power law holds.  
Reference~\cite{LIGOScientific:2025pvj} reported the slope of the two broken power laws to be $\alpha_1 = 1.7^{+1.2}
_{-1.8}$ and $\alpha_2 = 4.5^{+1.6}_
{-1.3}$, which are consistent with our inference, \defaultalphaone{} and \defaultalphatwo{} respectively.  
However, neither are consistent with the single power law index of the \fsnplp{} model, \plppowerlawindex{}.
Moreover, Ref.~\cite{LIGOScientific:2025pvj} argued that the high-mass population is not consistent with a single power law (plus a peak) as such a model overpredicts the rate at $m_1\sim 14\,M_{\odot}$.
In our case, however, \fsnplp{} is not ruled out, the critical difference being that we allow for a gap between the peak and the high-mass population.
Therefore, the presence of a gap in the model impacts the high masses and allows the entire high-mass population to be explained by a single power law.  
Observations of very high mass sources should help narrow down the high-mass power law, though the extent to which sources above $\sim 40\,M_{\odot}$ are formed from isolated channels is unclear, e.g.~\cite{Antonini:2025zzw,Roy:2025ktr,Tong:2025wpz,Antonini:2025ilj,Wang:2025nhf,Ray:2025xti,Sridhar:2025kvi}.

Quantitative model comparisons for models with such different parameter spaces and priors are subtle, especially when the priors are not particularly well-motivated. 
For example, the \fsnplp{} model has twice as many mass-ratio and spin parameters, all with uniform priors by default. 
With these well-known caveats, the Bayes Factor between the \fsnplp{} and the \default{} model is \plpoverdefaultbayesfactor, indicating   a mild preference for the latter.
At the same time, the ratio of maximum likelihood is \plpoverdefaultlikelihoodratio{}, indicating a substantial preference for the former model. 
A natural interpretation of this pattern is that the \fsnplp{} fits the data better, though at the expense of introducing additional parameters which have poorly constrained priors and therefore lead to lower marginal likelihoods. 
These additional parameters both add substantial prior volume and allow the model to fit the data better.

 However, the large maximum likelihood difference is not solely due to the extra degrees of freedom, but depends specifically on whether a gap is permitted. 
 We confirm this by repeating the analysis with the \fsnplp{} model, but now requiring $m^u_{\min}< 6\,M_{\odot}$ and $\delta m^u < 2\, M_{\odot}$, with all other prior bounds left unchanged. 
 This has the effect of forcing \emph{no gap} as $m^u_{\rm turn} < 8\,M_{\odot} < \mu^f$ and the primary mass rate remains nonzero for $m_1 \in (10, 20)\, M_{\odot}$. 
In that case, the maximum likelihood ratio drops to \plpforcednogapoverdefaultmaxlratio{}, a reduction of a factor of $\sim150$.  
The Bayes factor is also reduced to \plpforcednogapoverdefaultbayesfactor.  Therefore, the high relative likelihood is not \emph{solely} due to the additional degrees of freedom, but instead depends on whether the primary mass model permits a gap.

\section{The mass pairing model and its implications for the gap}
\label{app:plp-vs-pairing}

In this Appendix, we consider a population model which requires $m_1$ and $m_2$ to be drawn from the same population and then paired with some pairing function that depends only on the mass ratio.  
This is qualitatively similar to a model laid out in Ref.~\cite{Adamcewicz:2024jkr}, which found no evidence for a dip in the BBH merger rate near $10$--$15\, M_{\odot}$.
With this model, we confirm the conclusions of Ref.~\cite{Adamcewicz:2024jkr}, namely that the data are inconsistent with the same gap being present in both the $m_1$ and $m_2$ distributions.  

The mass model now contains a ``failed-SN" component (which we denote as ``$F$") and a ``higher-mass" component (which we denote as  ``$U$").
The components are 
\begin{align}
    p_{F}(m) &= \mathcal N(m; \mu, \sigma, m_{\min}, m_{\max})\,,\\
    p_{U}(m) &= p_{X}(m; \alpha_1 \dots \alpha_n)\,,
\end{align}
namely a Gaussian for the ``failed-SN" case and either a power law + peak or a broken power law for the ``higher-mass'' case. 
We restrict to the \plp{} model below.
The corresponding hyperparameters are defined by context, see App.~\ref{app:detailed-model-description}.
Instead of assigning each primary (and hence each binary) to one of these components, we here assign each individual BH independently.  
The model allows for each possible case of an object from either population pairing with an object of either population, and each case has its own spin and pairing distribution. 
Overall, we have four subpopulations:
\begin{align}
    p_{FF}(m_1, &m_2, \chieff) =  \nonumber\\
    &p_F(m_1) p_F(m_2)f(q|FF) p_{\chi}(\chieff|FF)\,,\label{eq:FF}\\ 
    p_{FU}(m_1, &m_2, \chieff) = \nonumber\\
    & p_F(m_1) p_U(m_2)f(q|FU) p_{\chi}(\chieff|FU)\,, \\
    p_{UF}(m_1, &m_2, \chieff) = \nonumber\\
    &p_U(m_1) p_F(m_2)f(q|UF) p_{\chi}(\chieff|UF)\,, \\
    p_{UU}(m_1, &m_2, \chieff) =\nonumber\\
    & p_U(m_1) p_U(m_2)f(q|UU) p_{\chi}(\chieff|UU)\label{eq:UU}\,,
\end{align}
where we have suppressed the hyperparameters. 
The mass ratio pairing is a power law and the spin distribution is a Gaussian. 
The overall population, that we refer to as the \pairing\, model, is the sum of components
\begin{align}
    p(m_1, &m_2, \chieff) \nonumber\\ 
    &= \lambda_{FF} p_{FF} + \lambda_{FU} p_{FU} + \lambda_{UF} p_{UF}+
    \lambda_{UU} p_{UU}\,,
    \label{eq:pairing-full}
\end{align}
where $\lambda_{FF},\lambda_{FU},\lambda_{UF},\lambda_{UU} \in (0, 1)$ and they sum to 1.  

\begin{figure*}
    \centering
    \includegraphics[width=0.99\textwidth]{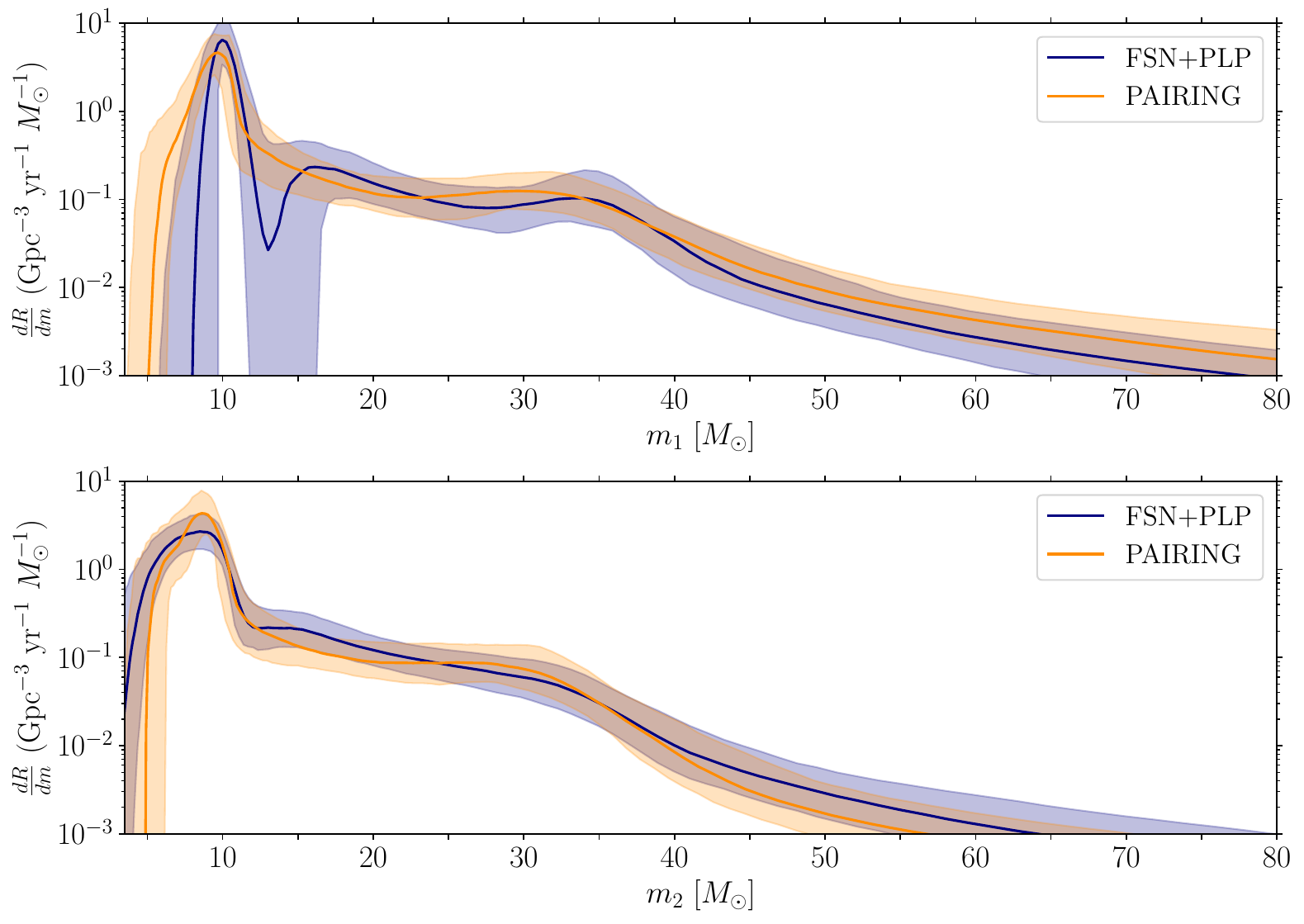}
    \caption{Same as Fig.~\ref{fig:compare-bpl-plp-fsn} but comparing the \fsnplp{} (blue) to the \pairing{} (yellow) model that assumes that the binary components have the same mass distribution as described in the text. }
    \label{fig:separate-pops-m1-m2-envelope}
\end{figure*}

The most pertinent feature of this model for our purposes is that the primary and secondary marginal mass distributions share the same support:  
\begin{equation}
    \label{eq:pairing-probability-rule}
    p(m_1) = 0 \iff p(m_2) = 0\,,
\end{equation}
where, e.g., 
\begin{equation}
    p(m_1) = \int p(m_1, m_2, \chieff)dm_2 d\chieff\,.
\end{equation}
This criterion is incompatible with most of the posterior of the \fsnplp{} model in Fig.~\ref{fig:compare-bpl-plp-fsn}, which shows zero  marginal rates for $m_1$ but not for $m_2$. 
Nonetheless, this criterion may be a desirable model feature for BHs formed independently and paired, e.g., under hierarchical or dynamical formation.  
For isolated binary evolution, however, different distributions of primary and secondary masses might be expected, even beyond what can be accommodated by the pairing function.  
More broadly, even if the physical distribution of component masses were identical, the ordering $m_1>m_2$ would result in distinct primary and secondary distributions either way, see Sec.~\ref{sec:conclusions}.

\begin{figure}
    \centering
    \includegraphics[width=0.5\textwidth]{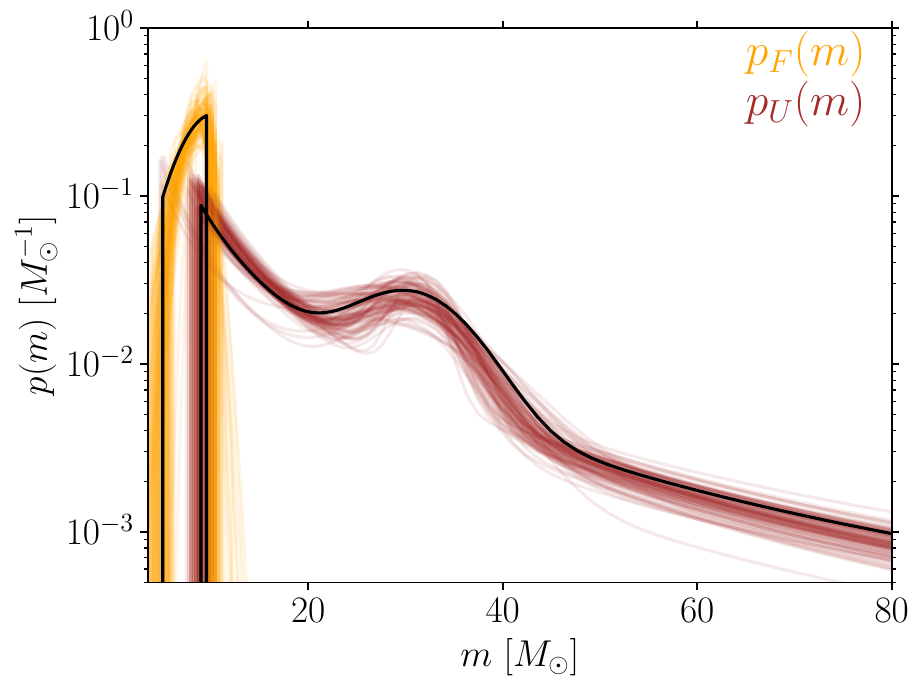}
    \includegraphics[width=0.5\textwidth]{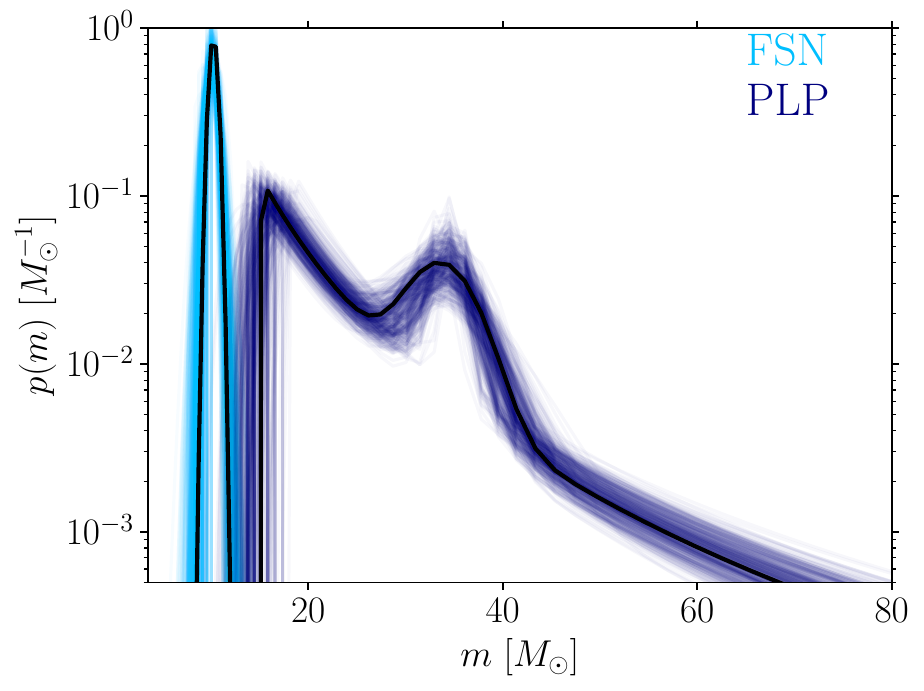}
    \caption{Maximum likelihood (black) and the 400 highest likelihood samples from the \pairing\, model (top) with $p_F(m)$ in orange and $p_U(m)$ in brown and the primary mass distribution of the \fsnplp{} model (bottom). }
    \label{fig:separate-pops-m-traces}
\end{figure}

We show the differential merger rates for the primary and secondary masses in Fig.~\ref{fig:separate-pops-m1-m2-envelope}, comparing the new \pairing\, model to the \fsnplp{} model from Fig.~\ref{fig:compare-bpl-plp-fsn}.
Under the \pairing\, model, the sharp minimum in the $m_1$ distribution just above $12\,M_{\odot}$ disappears; rather there is a broad local minimum for both components near $\sim 18-20\,M_{\odot}$. 
We do not find a gap in either these PPDs nor the highest likelihood hyperposterior samples, which we show in Fig.~\ref{fig:separate-pops-m-traces} for the \pairing\, model.  
This result is not surprising: under the \pairing\, model the support of the $m_1$ and $m_2$ distributions must coincide, e.g., Eq.~\eqref{eq:pairing-probability-rule}, but neither the \fsnplp{} nor \fsnbpl{} models have gaps in the $m_2$ distribution in Fig.~\ref{fig:compare-bpl-plp-fsn}. 

Direct model comparison via Bayes Factors is subject to the same interpretation concerns as discussed in App.~\ref{app:default-comparison}. 
Instead, to assess whether the additional parameters of the \pairing\, model lead to better fits (or contrarily if the constraints of Eq.~\eqref{eq:pairing-probability-rule} deteriorate the fit), we compare the 400 highest likelihood samples in Fig.~\ref{fig:separate-pops-m-traces}. 
The maximum likelihood ratio is \plpmaxlikelihoodoverseparatepops\, in favor of the \fsnplp{} model.\footnote{In the limit of an arbitrary number of subpopulations, the maximum likelihood value is achieved by a model that is a sum of delta functions~\cite{Payne:2022xan}. Maximum likelihood points need not reflect realistic astrophysical distributions under a finite amount of observations.} 
Under the \pairing{} model, the truncations of the $p_U(m)$ and $p_F(m)$ distributions overlap, leaving no space for a gap.

Under the \fsnplp{} model, there is instead a separation between the \fsn{} and high-mass populations, associated with a ``gap" in the distribution. 

\begin{figure}
    \centering
    \includegraphics[width=0.99\linewidth]{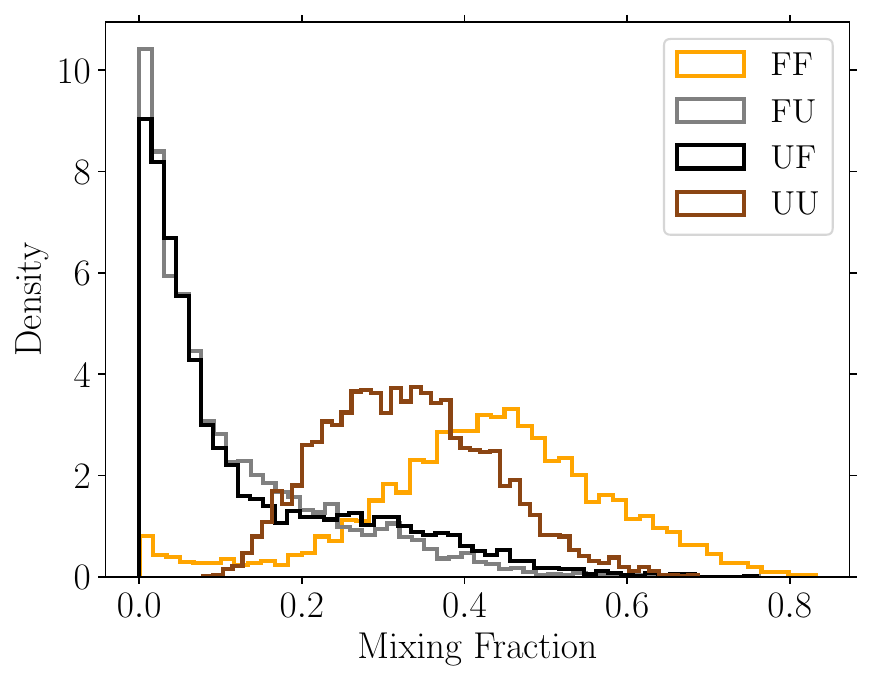}
    \caption{Marginalized posterior for the mixing fractions for the \pairing{} model defined in Eq.~\eqref{eq:pairing-full}. Labels mark the component of the population by the type of the primary and secondary, as described in the text and Eqs.~\eqref{eq:FF}-\eqref{eq:UU}. }
    \label{fig:pairing-mixing-fractions}
\end{figure}

Finally, we plot the inferred mixing fraction of each subpopulation per Eq.~\eqref{eq:pairing-full} in Fig.~\ref{fig:pairing-mixing-fractions}.  
The two most dominant contributions to the population are the $UU$ and $FF$ subpopulations making up $\sim 40\%$ and $\sim 30\%$ of the total population on average. 
The other components are consistent with zero, but with upper limits reaching $\sim 40\%$.  
There is therefore some preference for both binary components originating from the same subpopulation, with little mixing.
However, given the large model dimensionality and finite number of observations, the uncertainties are sizeable.

Nonetheless, the \pairing\, model does have certain desirable features in principle .  
For one, it explicitly models the $m_2$ distribution, although it is not independent of the $m_1$ one. 
This allows us to directly examine potential features in the $m_2$ distribution, such as the apparent presence of $\sim 10\,M_{\odot}$ secondaries merging with higher-mass primaries.\footnote{For example the events GW230605\_065343, GW230723\_101834, GW231114\_043211, GW231118\_005626~\cite{LIGOScientific:2025slb}.} 
Such events may be associated with a peak in $m_2$ but not in $m_1$, a scenario not explicitly modeled by our other approaches.   
We leave to future work extensions that independently model the $m_1$ and $m_2$ distributions, allowing for a gap in one but not necessarily the other.  
Such models have been used in Refs.~\cite{Antonini:2025zzw, Tong:2025wpz} to target a potential higher-mass gap due to pulsational pair-instability SNe.

\section{Alternative mass-ratio model of the \fsn{} subpopulation}
\label{app:other-mass-ratio}

In this Appendix, we consider a model that allows the \fsn{} subpopulation to have a preferred mass ratio (other than $q=1$ that is already allowed by a steep power law) in addition to having a preferred primary mass.  
This may be the case if, \emph{e.g.}, stable mass transfer impacts the binary component masses~\cite{vanSon:2022myr}.  

\begin{figure}
    \centering
    \includegraphics[width=0.5\textwidth]{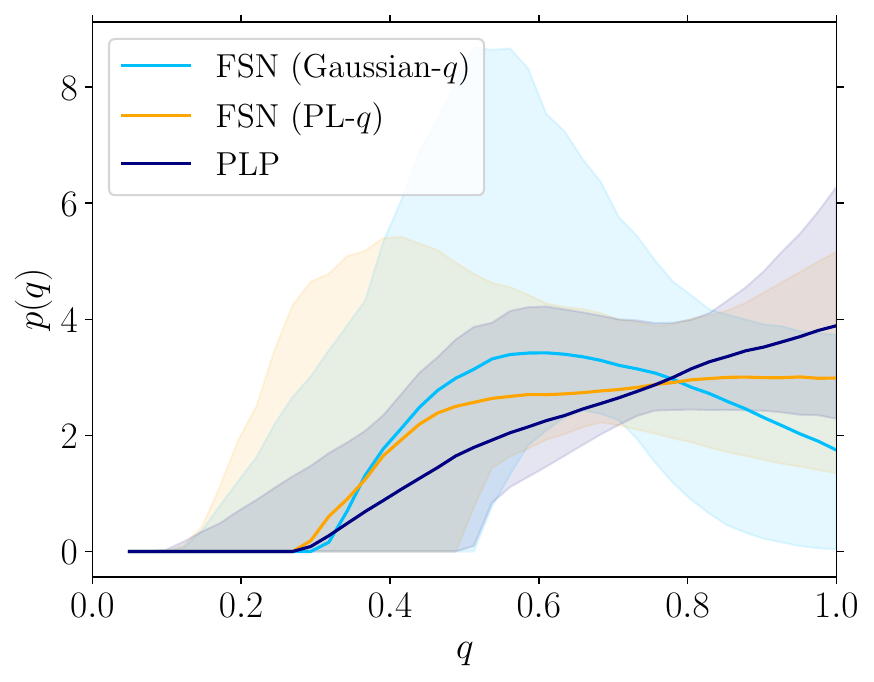}
    \caption{Same as the left panel of Fig.~\ref{fig:fsn-plp-q-chieff}, but treating the mass-ratio distribution for the \fsn{} component as a  Gaussian  (light blue).  We overplot for comparison results from Fig.~\ref{fig:fsn-plp-q-chieff} with a power law distribution in gold.}
    \label{fig:plp-otherq-mass-ratio}
\end{figure}
 
We therefore repeat the analysis using the \fsnplp{} model but replacing the mass ratio distribution of the \fsn{} subpopulation with a truncated Gaussian with mean, standard deviation, and truncation bounds as hyperparameters with uniform priors.  
We do not modify the mass ratio distribution of the high-mass subpopulation from a power law as we do not have a corresponding astrophysical interpretation for these binaries.
The mass ratio distributions under both models remain consistent, as shown in Fig.~\ref{fig:plp-otherq-mass-ratio}.  
The Gaussian distribution peaks at $q \sim 0.6$ and the Bayes Factor between the two models is $\sim 1$, indicating no preference.

\section{The impact of two low-mass outliers
}
\label{app:low-mass-sources}

\begin{figure}
    \centering
    \includegraphics[width=0.99\linewidth]{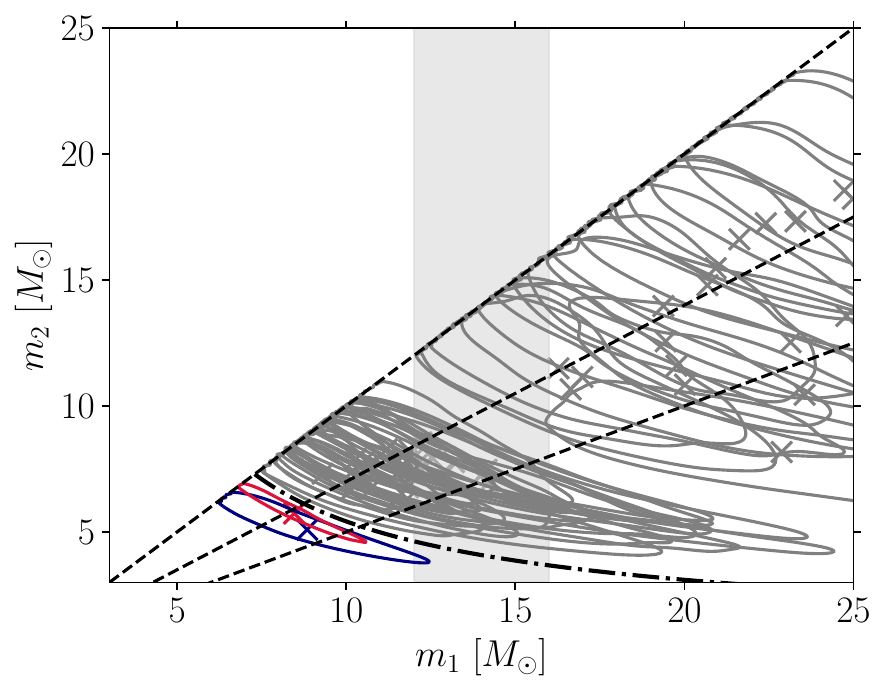}
    
    \caption{Similar to Fig.~\ref{fig:pop-informed-pe-nogap-vs-gap} but under the default parameter-estimation prior that is uniform in detector-frame component masses. The low-mass outliers GW190924 and GW230627 are highlighted in blue and red respectively, while other events are marked in gray.  We mark the chirp mass cut, $\mathcal{M}_c=6.35\,M_{\odot}$ in a black dash dot line}
    \label{fig:all-pe}
\end{figure}

Two BBHs have masses below the ``gap'' at $13\,M_{\odot}$, and yet do not appear to cluster with the other BBHs of the \fsn{} subpopulation, see Figs.~\ref{fig:pop-informed-pe-nogap-vs-gap} and~\ref{fig:all-pe}.
These are GW190924\_021846 (GW190924) and GW230627\_015337 (GW230627), with detector-frame chirp-masses $5.8^{+0.2}_{-0.2}\,M_{\odot}$ and $6.02^{+0.16}_
{-0.07}\,M_{\odot}$ respectively~\cite{KAGRA:2021vkt, LIGOScientific:2025slb}.  
Besides a low chirp mass, the events are otherwise unexceptional; both have $\chieff$ consistent with zero, and $q$ consistent with $1$ though with a large uncertainty allowing for $q\sim 0.5$ at $90\%$ credibility.
Despite the mass ratio uncertainty, the secondaries are almost certainly BHs, $m_2 \gtrsim 3.1\,M_{\odot}$~\cite{KAGRA:2021vkt, LIGOScientific:2025slb, Essick:2020ghc}.
This is in contrast to other sources with higher chirp masses but ambiguous secondaries: GW190814~\cite{LIGOScientific:2020zkf} with $m_2=2.6^{+0.1}_{-0.1}\,M_{\odot}$ and GW200210 with $m_2=2.8^{+0.4}_{-0.4}\,M_{\odot}$.
These events are typically excluded from BBH population analyses~\cite{LIGOScientific:2025pvj}.    

We consider the possibility that the outliers GW190924 and GW230627 also belong in a separate subpopulation of ``mass-gap" binaries with secondaries $\lesssim 5-6\,M_{\odot}$, together with GW190814 and GW200110, and perhaps GW230529~\cite{LIGOScientific:2024elc}.  
In practice, we consider only events with chirp mass $\mathcal M _c> 6.35\, M_{\odot}$ and $m_2 > 3.5\, M_{\odot}$, since there are no sources which straddle this divide~\cite{Essick:2023upv,  Golomb:2024lds} at 99\% credibility. 
We repeat our analysis with the model from App.~\ref{app:other-mass-ratio} which treats the mass ratio of the \fsn{} population as a truncated Gaussian, since the mass ratio inference is sensitive to these events.

\begin{figure*}
    \centering
    \includegraphics[width=0.99\textwidth]{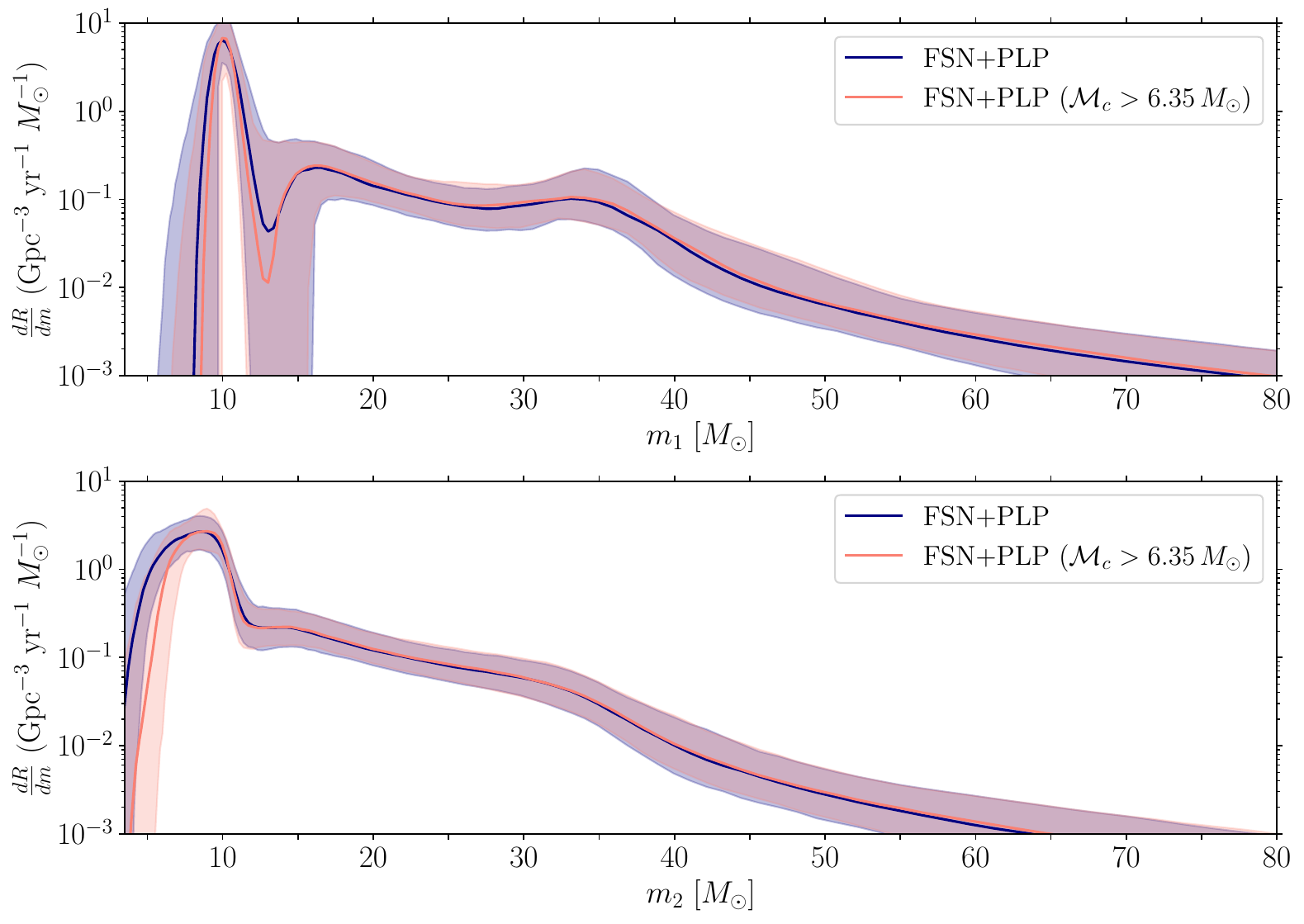}
    \caption{Same as Fig.~\ref{fig:compare-bpl-plp-fsn} but comparing the \fsnplp{} (blue) result from App.~\ref{app:other-mass-ratio} to the same model but only inferring the population which has a source-frame chirp mass $\mathcal M_c > 6.35\,M_{\odot}$, effectively removing the sources GW190924 and GW230627 (pink). In both cases, the mass ratio of the \fsn{} subpopulation is modeled with a truncated Gaussian.}
    \label{fig:plp-otherq-m1-m2-no-low-mass}
\end{figure*}

\begin{figure}
    \centering
    \includegraphics[width=0.5\textwidth]{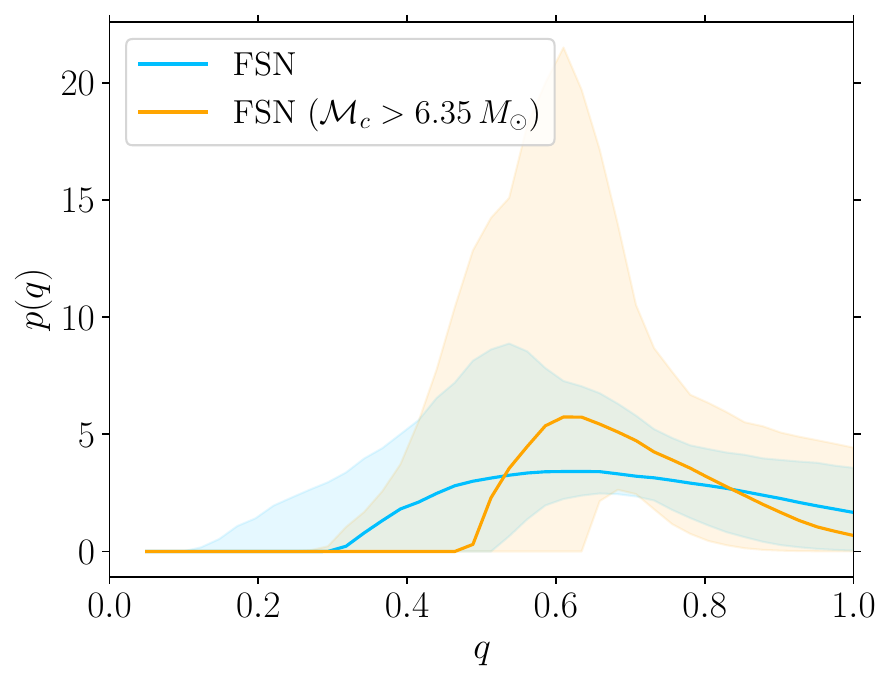}
    \caption{Effect of the low-mass outliers on the inferred mass ratio distribution. We plot the \fsn{} distribution from the \fsnplp{} Gaussian-$q$ analysis of Fig.~\ref{fig:plp-otherq-mass-ratio} in light blue and the same result after excluding the sources GW190924 and GW230627 (orange). }
    \label{fig:plp-otherq-mass-ratio-no-low-mass}
\end{figure}

We show the impact of the two low-mass outliers on the differential merger rate in Fig.~\ref{fig:plp-otherq-m1-m2-no-low-mass}.  
The primary effect of the additional cuts (and therefore removing the outlier events) is a sharper \fsn{} peak with $\sigma^f=$\plpnolowmassfsnwidth (vs. \plpfsnbumpwidth\, previously) and an increased preference for a gap.
Now \result{93\%} of the posterior supports the presence of a gap.  
The fact that excluding sources below the \fsn{} peak leads to rate changes above the peak is an outcome of the model choice, and especially the tightening of the \fsn{} Gaussian.
Though since all rates stay consistent within uncertainties, the headline conclusions about the existence of the gap are robust.
Figure~\ref{fig:plp-otherq-mass-ratio-no-low-mass} shows the impact of the low-mass outliers specifically on the inferred \fsn{} mass ratio distribution.
Excluding them leads to an even more peaked distribution at $q\sim 0.6-0.7$ and with a higher cutoff of \plpnolowmassqmin{}, compared to \plpfailedqmin{} previously.
However within 90\% uncertainties, the distributions are consistent.

\new{
\section{Allowing for structure in the component spin distributions}
\label{app:different-spin-distribution}

In this Appendix, we explore the impact of the spin model on our main results. 
We repeat the \fsnplp{} headline inference, but with a $\chieff$ distribution that is a sum of two Gaussians for both the \fsn{} and \plp{} populations.  
Such a distribution allows for asymmetry in the $\chieff$ distribution for each component, and also for potential substructure in the spin distribution resulting from  different formation channels. 

We use agnostic priors on the mean and standard deviation of the two component gaussians $\mu^{\bullet}_{1,\chi}, \mu^{\bullet}_{2, \chi}\sim U(-1, 1)$, $\sigma^{\bullet}_{1,\chi}, \sigma^{\bullet}_{2, \chi}$, where the $\bullet$ indicates that this holds for both the \fsn{} and \plp{} subpopulations. 
In Fig.~\ref{fig:other-spin-chieff}, we show the inferred $\chieff$ distribution for both subpopulations. 
The \fsn{} spin distribution is minimally affected, while the \plp{} one is now more narrowly concentrated around zero. 
Though the two distributions more more similar than those of Fig.~\ref{fig:fsn-plp-q-chieff} (right panel), they remain distinct.  
In particular, the \plp{} distribution is shifted to lower values, with essentially no support at high positive spin.  

\begin{figure}
    \centering
    \includegraphics[width=0.99\linewidth]{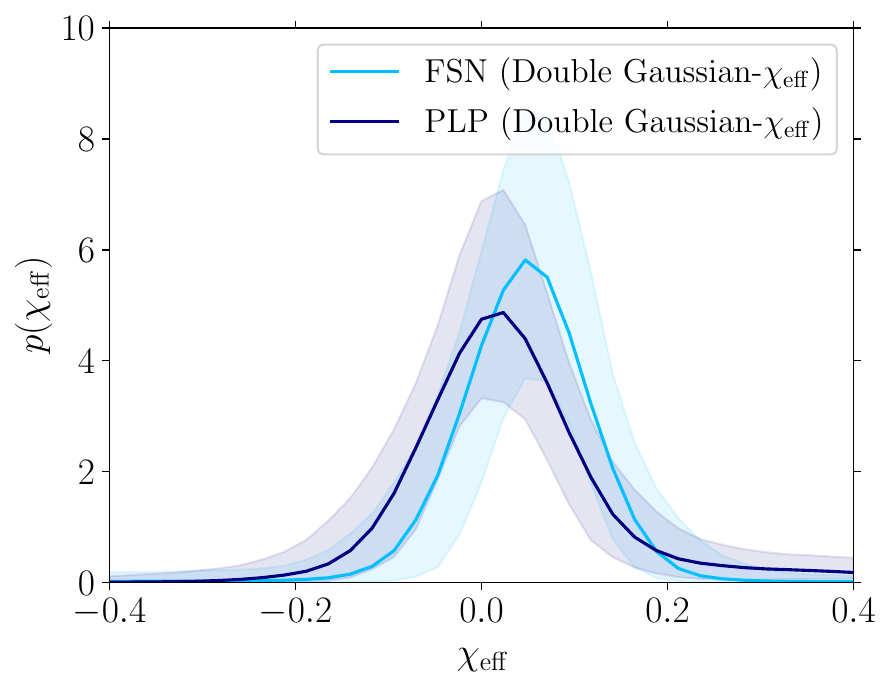}
    \caption{Same as the right panel of Fig.~\ref{fig:fsn-plp-q-chieff}, but using a $\chieff$ model that is a sum of two Gaussian components rather than a single Gaussian.}
    \label{fig:other-spin-chieff}
\end{figure}
}

\section{Justification for the effective zero-rate estimate}
\label{app:zero-rate-justification}

In this Appendix, we expand upon the heuristic calculation for the ``effectively zero" rate we use in Sec.~\ref{sec:failed-sn-results}.  
We first emphasize that the rate estimate of $10^{-3}$\rateunits{} can only be considered a ``zero rate" in the region of mass space we are interested, $\sim 15 \,M_{\odot}$.  Sensitivity to higher mass sources implies that even substantially lower rates can lead to detections at higher masses~\cite{LIGOScientific:2025pvj,LIGOScientific:2025slb}.  

The calculation proceeds as follows. 
From Ref.~\cite{LIGOScientific:2025slb}, we use an estimate for the O4a sensitive volume time ($VT$) for $20 + 20\,M_{\odot}$ mergers of $9.6\, {\rm Gpc^3 yr}$, see \emph{e.g.} Fig.~1 top panel. 
Precisely estimating the rate at $15+15\,M_{\odot}$ would require additional sensitivity simulations~\cite{LIGOScientific:2025hdt}. 
But given that we are only seeking a coarse estimate of what can be considered a ``small rate," we conservatively adopt the same $VT$ at $15+15\,M_{\odot}$ as $20+20\,M_{\odot}$.  
Since sensitivity is roughly proportional to chirp mass which scales with the total mass, we are therefore overestimating the probed $VT$ by about $(20/15)^3\approx2$.  
To estimate the total $VT$ probed over the entire duration of O1-O4a, we compare the \emph{total} O4a $VT$ ($5.28\times 10^{-3}\,{\rm Gpc^3yr}$) and that of the combined O3 ($3.21\times10^{-3}\,{\rm Gpc^3 yr}$) analysis~\cite{LIGOScientific:2025hdt}.  
This comparison suggests that at the level of approximation we consider, jointly O1-O4a featured at most double the observed $VT$ of O4a alone. 
So again conservatively, we take the observed volume time over O1-O4a in the relevant mass range to be
\begin{equation}
VT_{15\,M_{\odot}}\approx  2 \times 9.8\, {\rm Gpc^3yr} \approx 20  \, {\rm Gpc^3yr}\,.  
\end{equation}

In the main text, we then proceed to estimate that for a mass range of $\sim 5\,M_{\odot}$ around $15\,M_{\odot}$, we would expect to observe $\approx R\times 5\,M_{\odot} \times VT_{15\,M_{\odot}}$ sources where $R$ is the relevant differential merger rate.  
At a rate of $10^{-3}$\,\rateunits{} this works out to $\approx 0.1$ total detections.
Since observing $0$ source is consistent with an expectation of $0.1$, we consider the rate of $10^{-3}$\,\rateunits{} to be indistinguishable from zero at current sensitivity.
This is certainly an overestimate due to the choices we make here, the true value may be nearly an order of magnitude lower.

\bibliography{references}

\end{document}